\documentclass[12pt]{iopart}
\usepackage{iopams}                            
\usepackage{subfigure}
\usepackage{graphicx}
\usepackage{color}
\usepackage{subeqn}                            
\usepackage[en-US]{datetime2}                         
\newcommand{\Schrodinger}{Schr{\"o}dinger}     
\newcommand{\ansatz}{ans{\"a}tz}               
\newcommand{\PT}{\mathcal{PT}}                 
\newcommand{\calP}{\mathcal{P}}
\newcommand{\calT}{\mathcal{T}}
\newcommand{\calK}{\mathcal{K}}

\newcommand{\ef}[1]{(\ref{#1})}                
\newcommand{\dd}[1]{\,\mathrm{d}#1\,}

\newcommand{\sech}{\,\mathrm{sech}}
\newcommand{\csch}{\,\mathrm{csch}}
\newcommand{\Imag}[1]{ \,\mathrm{Im}{\lbrace \,#1\,\rbrace}}   
\newcommand{\tint}{\!\int\!}                   
\newcommand{\pdv}[2]{\frac{\partial #1}{\partial #2}}
\newcommand{\qc}{\>,\quad}

\newcommand{\notag}{\nonumber}                 

\newcommand{\bq}{\begin{equation}}
\newcommand{\be}{\begin{equation}}
\newcommand{\ee}{\end{equation}}
\newcommand{\bqa}{\begin{eqnarray}}
\newcommand{\eea}{\end{eqnarray}}

\newcommand{\ba}{\begin{eqnarray}}
\newcommand{\eq}{\end{equation}}
\newcommand{\ea}{\end{eqnarray}}
\eqnobysec                                     
\begin{document}

\title[Anti-$\PT$ stability]
{Behavior of solitary waves of coupled nonlinear \Schrodinger\ equations 
subjected to complex external periodic potentials with anti-$\PT$ symmetry}%

\author{
   Efstathios G. Charalampidis$^1$,
   Fred Cooper$^{2,3}$,
   John F. Dawson$^4$, 
   Avinash Khare$^5$, 
   Avadh Saxena$^3$}
\address{$^1$Department of Mathematics, California Polytechnic State University, San Luis Obispo, CA 93407-0403, United States of America}
\address{$^2$The Santa Fe Institute, 1399 Hyde Park Road, Santa Fe, NM 87501, United States of America}
\address{$^3$Theoretical Division, Los Alamos National Laboratory, Los Alamos, NM 87545, United States of America}
\address{$^4$Department of Physics, University of New Hampshire, Durham, NH 03824, United States of America}
\address{$^5$Physics Department, Savitribai Phule Pune University, Pune 411007, India}
\ead{echarala@calpoly.edu}
\ead{cooper@santafe.edu}
\ead{john.dawson@unh.edu}
\ead{khare@physics.unipune.ac.in}
\ead{avadh@lanl.gov}
\vspace{4pt}
\begin{indented}
\item[]\DTMnow
\end{indented}
\begin{abstract}
We discuss the response of both moving and trapped solitary wave solutions of a 
nonlinear two-component nonlinear \Schrodinger\  system in 1+1 dimensions to an anti-$\PT$ 
external periodic complex potential. The dynamical behavior of perturbed solitary waves is
explored by conducting numerical simulations of the nonlinear system and using a collective
coordinate  variational approximation. We present case examples corresponding to
choices of the parameters and initial conditions involved therein.
The results of the collective coordinate approximation are compared against
numerical simulations where we observe qualitatively good agreement between 
the two.
Unlike the case for a single-component solitary wave in a complex periodic
$\PT$-symmetric potential, the collective coordinate equations do not have 
a small oscillation regime, and initially the height of the two components 
changes in opposite directions often causing instability. We find that the 
dynamic stability criteria we have used in the one-component case is proven 
to be a good indicator for the onset of dynamic instabilities in the present
setup.

\end{abstract}
\submitto{\jpa}
\vspace{1pc}
\noindent{\it Keywords}: Collective coordinates, variational approach, dissipation functional, spatio-temporal dynamics.
\maketitle
%
\section{\label{s:Intro}Introduction}

%
The study of solitons in nonlinear partial differential equations (PDEs) with 
non-Hermitian potentials is an important and growing area of research. Specifically,
such parity-time or $\PT$-symmetric PDEs have been studied in detail \cite{BenderBook}. 
Subsequent to 
the introduction of $\PT$-symmetry~\cite{Bender1998} and the ensuing intense research
on this topic for a decade and half, the concept of anti-$\PT$ symmetry was first 
introduced in the context of optics~\cite{Ge2013} by appropriately arranging the 
effective optical potential spatially. For anti-$\PT$-symmetric systems, one has 
the $\PT$ operator commuting with the Hamiltonian $[H, \PT ]=0$, but in addition 
$(\PT) ^2 = -1$ in contrast with the $+1$ as for the $\PT$-symmetric systems.
Its implementation requires the introduction of at least two components in the 
wave function.
Recently, there have been several realizations of the anti-$\PT$ symmetry 
such as in coupled atom beams~\cite{Peng2016}, optical waveguides with imaginary 
couplings~\cite{Zhang2019}, electrical circuit resonators~\cite{Choi2018}, as 
well as cold atom based optical four-wave mixing~\cite{Jiang2019}. Moreover, 
besides optical systems with constant refraction~\cite{Yang2017}, many other 
experiments have realized the anti-$\PT$ symmetry in atomic \cite{Wang2016, Chuang2018} 
and optical \cite{Konotop2018, qLi2019, Ke2019} systems. In addition, there are several
other applications that are related to waveguide arrays~\cite{Ke2019}, spin chains~\cite{Couvreur2017}, 
phase transitions~\cite{Lee2019}, diffusive systems~\cite{Li2019}, information 
flow~\cite{Chakraborty2019} as well as non-Markovian processes~\cite{Haseli2014}. 

%
Recently, anti-$\PT$ symmetric couplers have been analyzed by Konotop and 
Zezyulin~\cite{PhysRevA.99.013823} which lead to lasing and coherent perfect
absorption. These systems are experimentally reproduced by having two waveguides 
locally coupled through an anti-$\PT$ symmetric medium. Here we generalize the
treatment of Konotop and Zezyulin to the case of coupled nonlinear \Schrodinger\ 
equations (NLSEs) which are 
individually subjected to external potentials as well as coupled by an antisymmetric 
medium. Such systems can be produced in nonlinear optics in the 
wave-guiding approximation.

In order to implement anti-$\PT$ symmetry in the NLSE one requires at least a  
two-component NLSE.  Previously we studied exact  trapped solitary wave solutions
of the two-component NLSE in an external complex supersymmetric potential which had 
$\PT$ symmetry \cite{trappedPT}.  In that situation we found regions of stability and 
instability predicted by both the small oscillation equations for the collective 
coordinates (CCs), as well as the dynamic criteria and a systematic numerical stability analysis. 
Although we were able to find exact solutions of  the two-component NLSE in some external 
 complex supersymmetric potentials similar to those considered in  \cite{trappedPT} but having  anti-$\PT$ symmetry, all the solutions 
we have found so far are unstable.  To better understand the behavior and stability of  solitons  in the two-component NLSE  in complex external potentials having anti-$\PT$ symmetry, we will study here the simpler question of what happens to stable  solitary wave solution of the two-component NLSE when then subjected to an external complex periodic potential having anti-$\PT$ symmetry.
 This generalizes 
a previous problem in~\cite{PhysRevE.94.032213} 
that we studied for the single-component NLSE solitary wave which was placed in a complex 
periodic external potential with $\PT$ symmetry. 

In the present work, the NLSE soliton solution is a solution of a 
two-component NLSE. These two-component solitons (individually identified as $\psi_{1}$
and $\psi_{2}$ hereafter) have the property that they have anti-$\PT$ 
symmetry, which itself requires that $\psi_2(x,t) = \kappa \, \rmi \, \psi_1(-x,-t)$ with
$\kappa = \pm 1$.  Also for $\PT$ symmetry, the single-component complex external potential 
we chose previously was of the form $V(x) = a_1 \cos k_1 x + \rmi \, a_2 \sin k_2x$. For 
anti-$\PT$ symmetry, the complex matrix potential, $U(x)$ takes the form $U(x) = \sigma_{0}V_0(x)+ \rmi \, \sigma_3 V_1(x) +\rmi \, \sigma_1 W(x)$, 
where $V_0=  a_1 \cos k_1 x $, $ V_1(x) =a_2 \cos k_2 x$, $W(x) =  a_3 \cos k_3 x$, 
with $\sigma_{0}$ the $2\times2$ identity matrix, and $\sigma_{1}$ and $\sigma_{3}$ being the Pauli matrices.
The second term in the potential changes two things qualitatively. Firstly, calling the 
initial position of the center of the soliton $q_0$, 
we have that $V_{1}(x)$ has a minimum  at $k_2 q_0=0$ and a maximum at  $k_2 q_0= \pi$ with magnitude $a_2$.
We will show that this prevents us from obtaining
a small oscillation expansion for the CC 
approximation.  Related to this,
 the effect of the $\sigma_3$ term in the potential is to cause  
$\psi_{1}$ and $\psi_{2}$ 
to initially grow and decay linearly in time, respectively 
(or vice versa depending on $q_0$). This is the main 
reason for the fact that when $a_2 \neq 0$, the soliton becomes dynamically unstable whether 
it is trapped or moving. In spite of this, when the soliton experiences the external potential, 
the widths  of both components remain almost identical. This is true also for the position and
momentum of each component. The complex potential $\rmi \, \sigma_1 a_3 \cos k_3 x$ connects 
directly the two components of the wave function. Having $a_3$ small and nonzero has minimal 
effect on changing the ``mass'' $M_i$ of the two components if $a_2$ is zero. (Here we define 
$M_i := \int \psi_i^* \psi_i dx$ for each component.) 
We show that a CC description of the two-component wave function describes reasonably well
the response of the solitary wave to this anti-$\PT$ external potential if we allow   
the masses and phases of the two components to differ, but keep  the  position, momentum, 
width and ``chirp'' to be the same for both components.

In particular, when $a_3=0$, we chose the strength of the two external potentials to  
match those we used in our single-component case~\cite{PhysRevE.94.032213}.  We also 
investigated the ability of the dynamical indicator of instability, i.e., whether 
$dp(t)/ d v (t)$ becomes negative~\cite{PhysRevE.84.026614}, to indicate dynamical 
instability for this two-component NLSE system. Here $q(t)$ and $v(t)$ are canonical 
variables with $v(t)=\dot{q}(t)$, and $p(t)$ is related to the average value of the 
momentum operator $ -i \partial_x$.
This indicator visually shows the instability near where $p(t)$ is turning around 
from a maximum or a minimum. 
Since $ \sigma_3 V_1(x)$ initially 
places the two components in opposite directions, it 
is the major cause for all the various ways that the initial solitary wave can go unstable. 
These phenomena are reasonably well captured by the eight collective coordinate 
(8CC) approximation which is compared with direct numerical simulations.  

The structure of this paper is as follows.  In section~\ref{s:antiPT} we discuss the conditions that anti-$\PT$ symmetry places on the wave function and the external potential. In section~\ref{s:exactsol} we obtain exact moving anti-$\PT$ symmetric solutions of the two-component NLSE and use Derrick's theorem to show that they are stable to scale transformations.  In section~\ref{s:collective} we review the CC 
formalism and in section~\ref{s:8CC} we introduce our choice of 8CCs, partially motivated by the numerical simulations.  In section~\ref{s:comp} we show how to compare the results of the numerical simulation with the time evolution of the CC
by relating the 
 CCs to low order moments of the numerically determined wave function.  In section~\ref{ss:typical} we give some typical behaviors for different values of the parameters describing the complex external potential.  In section~\ref{s:pvcurve} we discuss the stability criterion $dp/dv <0$ and show that in all the cases we study both the 8CC and numerical determinations of $p(v)$, it leads to the conclusion that these solitary waves are dynamically unstable. 
 In section~\ref{s:conclusions} we state our conclusions and present directions for future study. 
%
\section{\label{s:antiPT}Anti-$\PT$ systems}

In the present work, we consider a two-component nonlinear \Schrodinger\ equation 
(NLSE) in 1+1 dimensions:
\begin{equation}\label{AntiPTeq}
   \bigl \{\,
      \rmi \, \partial_t
      +
      \partial_{x}^2 
      +
      g \, [\, \Psi^{\dag}(x,t) \Psi(x,t) \,]
      -
      U(x) \, \} \,
      \Psi(x,t)
      =
      0 \>,
\end{equation}
where
\begin{equation}\label{Psidef}
   \Psi(x,t)
   =
   \Bigl ( \begin{array}{c}
      \psi_1(x,t) \\
      \psi_2(x,t)
   \end{array} \Bigr ) \in\mathbb{C}^{2}\>
\end{equation}
is the wave function and $g$ the nonlinearity strength.  Here  
$x$ and $t$ stand for the spatial and temporal variables,
respectively, and subscripts in Eq.~\ef{AntiPTeq} for differentiation
with respect to the variables highlighted therein (unless stated
otherwise). The matrix function $U(x)$ is the external potential 
that we describe next.

For two-component systems, the space ($\calP$) and time ($\calT$) 
reversal operators are defined by:
\begin{subeqnarray}\label{PTops}
   \calP \, \Psi(x,t)
   &=&
   \Psi(-x,t) \>,
   \label{Pop} \\
   \calT \Psi(x,t)
   &=& 
   \rmi \, \sigma_2 \, \calK \, \Psi(x,-t)
   \label{Top} \\
   &=&
   \Bigl ( \begin{array}{cc} 
      0 & \calK \\
      - \calK & 0
   \end{array} \Bigr ) \,
   \Bigl ( \begin{array}{c}
      \psi_1(x,-t) \\
      \psi_2(x,-t)
   \end{array} \Bigr )
   =
   \Bigl ( \begin{array}{c}
      \psi_2^{\ast}(x,-t) \\
      - \psi_1^{\ast}(x,-t)
   \end{array} \Bigr ) \>, 
   \notag 
\end{subeqnarray}
where $\calK$ is the complex conjugate operator with the property $\calK^2 = 1$. 
The parity and time-reversal operations commute, i.e., $[\,\calP,\calT \,] = 0$, 
and obey the relations $\calP^2 = 1$ and $\calT^2 = -1$, so that $(\PT)^2 = -1$. 
Then the $\PT$ operation on $\Psi$ is given by
\begin{equation}\label{PTPsi}
   \PT \Psi(x,t)
   =
   \Bigl ( \begin{array}{c}
      \psi_2^{\ast}(-x,-t) \\
      - \psi_1^{\ast}(-x,-t)
   \end{array} \Bigr ) \>.
\end{equation}
For anti-$\PT$ symmetry, the linear part of Eq.~\ef{AntiPTeq} must commute with the 
$\PT$ operator
\begin{equation}\label{PTUcommute}
   [\, \PT \,,U(x) \, ] = 0
   \>\quad\Rightarrow \quad
   U(x)
   =
   \PT \,U(x) \, (\PT)^{-1} \>.
\end{equation}
Let 
\begin{equation}\label{Ugendef}
   U(x)
   =
   \Bigl ( \begin{array}{cc}
      U_0(x) & V_2(x) \\
      V_2(x) & U_1(x)
   \end{array} \Bigr ) \in\mathbb{C}^{2\times 2}\>,
\end{equation}
be dependent on $x$ only. 
Then, \ef{PTUcommute} requires that
\begin{eqnarray}
   \Bigl ( \begin{array}{cc}
      U_0(x) & V_2(x) \\
      V_2(x) & U_1(x)
   \end{array} \Bigr )    
   &=
   \Bigl ( \begin{array}{cc} 
      0 & \calK \\
      - \calK & 0
   \end{array} \Bigr ) \,
   \Bigl ( \begin{array}{cc}
      U_0(-x) & V_2(-x) \\
      V_2(-x) & U_1(-x)
   \end{array} \Bigr )    
   \Bigl ( \begin{array}{cc} 
      0 & - \calK \\
      \calK & 0
   \end{array} \Bigr ) 
   \notag \\
   &=
   \Bigl ( \begin{array}{cc}
      \calK U_1(-x) \calK & -\calK V_2(-x) \calK \\
      - \calK V_2(-x) \calK & \calK U_0(-x) \calK
   \end{array} \Bigr ) \>,
   \label{UpaUpa}
\end{eqnarray}
from which we conclude that
\begin{equation}\label{U0U1V2conditions}
   U_0(x) = U_1^{\ast}(-x)
   \qc
   V_2(x) = - V_2^{\ast}(-x) \>.
\end{equation}
Setting $U_0(x) = V_0(x) + \rmi \, V_1(x)$ with $V_0(x),V_{1}(x)\in\mathbb{R}$, we find that
$U_1(x) = U_0^{\ast}(-x) = V_0(-x) - \rmi \, V_1(-x)$. 
This way, we can write Eq.~\ef{Ugendef} as
\begin{equation}\label{UgenII}
   U(x)
   =
   \Bigl ( \begin{array}{cc}
      V_0(x) + \rmi \, V_1(x) & V_2(x) \\
      V_2(x) & V_0(-x) - \rmi \, V_1(-x)
   \end{array} \Bigr ) \>.
\end{equation}
If 
we additionally require that $V_0(-x) = V_0(x)$ and $V_1(-x) = V_1(x)$, 
i.e., $V_{0},V_{1}$ are even functions, and 
$V_2(x) := \rmi \, W(x)$ with $W(x)\in\mathbb{R}$ and even, then $U(x)$ is now 
given by
\begin{eqnarray}\label{Udef}
   U(x)
   &=
   \Bigl ( \begin{array}{cc}
      V(x) & \rmi \, W(x) \\
      \rmi \, W(x) & V^{\ast}(x)
   \end{array} \Bigr ) 
   \qc
   V(x):= V_0(x) + \rmi \, V_1(x) \>.
\end{eqnarray}
It will be useful to split $U(x)$ into real and imaginary parts 
via $U(x) := U_0(x) + \rmi \, U_1(x)$, where
\begin{equation}\label{U0U1def}
   U_0(x)
   =
   \Bigl ( \begin{array}{cc}
      V_0(x) & 0 \\
      0 & V_0(x)
   \end{array} \Bigr ) 
   \qc
   U_1(x)
   =
   \Bigl ( \begin{array}{cc}
      V_1(x) & W(x) \\
      W(x) & -V_1(x)
   \end{array} \Bigr ) \>.   
\end{equation}
Calling $\sigma_0=\mathbb{I}_{2}$, i.e., the $2\times 2$ unit matrix, we can write 
\bq U(x) = \sigma_0 V_0(x) + i \sigma_3 V_1(x) + i \sigma_1 W(x).
\eq
Eigenstates of the anti-$\PT$ operator satisfy the equation
\begin{equation}\label{PTeigen}
   \PT \, \Psi_{\kappa}(x,t)
   =
   \kappa \, \rmi \, \Psi_{\kappa}(x,t)
   \qc
   \kappa = \pm 1 \>,
\end{equation}
from which we conclude that the components satisfy:
\begin{equation}\label{PSI12eigen}
   \psi_{2\,\kappa}(x,t)
   =
   \kappa \, \rmi \, \psi_{1\,\kappa}^{\ast}(-x,-t) \>.\
\end{equation}

%
\section{\label{s:exactsol}Exact solitary wave solutions when $U(x)\equiv 0$}

In the absence of external potentials, 
Eq.~\ef{AntiPTeq} reduces to:
\begin{subeqnarray} \label{PsiEq}
   \{\,
      \rmi \,
      \partial_t
      +  
      \partial_x^2
      + 
      g \, ( | \psi_1(x,t) |^{2}+ | \psi_2(x,t) |^{2}) \,
   \} \, \psi_1(x,t) 
   &=
   0 \>,
   \label{PsiEq-a} \\
   \{\,
      \rmi \,
      \partial_t
      +  
      \partial_x^2
      +
      g \, ( | \psi_1(x,t) |^{2}+ | \psi_2(x,t) |^{2}) \,
   \} \, \psi_2(x,t) 
   &= 0 \>,
   \label{PsiEq-b}
\end{subeqnarray}
whence it is easy to show that the traveling solitary wave solution
\begin{subeqnarray}\label{A-Psi}
   \psi_1(x,t)
   &=
   A_1 \, \beta  \sech[\, \beta \, (x-vt) \,] \, 
   \exp \{\, \rmi \, [\, p \, (x-v t) - \theta(t)  \,] \, \} \>,
   \label{A-Psi-a} \\
   \psi_2(x,t)
   &=
   A_2 \, \beta  \sech[\, \beta \, ( x -v t) \,] \, 
   \exp \{\, \rmi \, [\, p \, (x-v t) - \theta(t)  \,] \, \} \>,
   \label{A-Psi-b}
\end{subeqnarray}
with real frequencies is an exact solution of \ef{PsiEq} provided that
\begin{equation}\label{balphaconditions}
   g \, (\, |A_1|^2 + |A_2|^2 \,)
   =
   2  \>, 
   \quad
   p = \frac{v}{2} \>, 
   \quad 
   \theta(t) = - (\, p^2 + \beta^2 \,) \, t \>.
\end{equation}
These solutions are eigenstates of the anti-$\PT$ operator $\forall x,t$ if 
\begin{equation}\label{A-antiPt}
   A_2
   =
   \rmi \, \kappa \,  A_1^{\ast}
   \qc
   \kappa
   =
   \pm 1 \>,
\end{equation}
in which case, if we set $A_1 = A$ and $A_2 =  \rmi \, \kappa \, A$, then $A^2 = 1/g$.
Normalization integrals are given by:
\begin{eqnarray}\label{staticM}
   M_1 = \tint \dd{x} | \psi_1(x,t) |^2 = 2 \beta A_1^2
   \qc
   M_2 = \tint \dd{x} | \psi_2(x,t) |^2 = 2 \beta A_2^2 \>,
\end{eqnarray}
such that the condition in Eq.~\ef{balphaconditions} becomes $g \, (M_1 + M_2) = 4 \beta$.
Given now this form of the exact solution, the self-interaction potential 
term commutes with the $\PT$ operator $\forall t$. For the soliton at rest, 
Eq.~\ef{A-Psi} reduces to
\begin{subeqnarray}\label{staticsol}
   \psi_1(x,t)
   &=
   A \, \beta  \sech( \beta \, x ) \, 
   \exp \{\, \rmi \, (- \beta^2 t \,) \, \} \>,
   \label{staticsol-a} \\
   \psi_2(x,t)
   &=
   A \, \beta  \sech( \beta \, x ) \, 
   \exp \{\, \rmi \, ( - \beta^2 t \pm  \pi/2  \,) \, \} \>.
   \label{staticsol-b}   
\end{subeqnarray}

%
\subsection{\label{ss:Derrick}Derrick's theorem}

We can use the scaling argument of Derrick \cite{doi:10.1063/1.1704233} to determine if
the two-component static solutions of \ef{staticsol} are stable to scale transformations. 
For the sake of completeness in the present discussion, we introduce the nonlinearity 
exponent, identified as $k$ hereafter, which allows us to show that the stability depends 
on $k$. For the single-component NLSE at hand, the solutions are unstable to either 
blowup or collapse when $k >2$~\cite{Cooper:2017aa}. Here we will confirm that the exact 
solutions we found for $k=1$ are stable to scale transformations. 
To that effect, let us recall first the Hamiltonian given itself by 
\begin{equation}
   H 
   = 
   \tint \dd{x} 
   \Big\{\,
      \frac{1}{2} \, |\,\partial_x \Psi(x) \,|^2
      -
      \frac{g}{k+1} \, [\, \Psi^{\dag}(x) \Psi(x) \,]^{k+1} \,
   \Bigr \} \>,
\end{equation}
where $\Psi(x)$ 
denotes the static two-component solution of \ef{PsiEq}.
It is well known that using stability with respect to scale transformation to 
understand domains of stability applies to this type of Hamiltonian. 
If we make the scale transformation of the solution of the form
\begin{equation}\label{scaletrans}
   \Psi(x) \mapsto \alpha^{1/2} \Psi(\alpha x) \>=\alpha^{1/2} \Psi(y), \,\,y:=\alpha x
\end{equation}
which preserves the normalization, i.e., $M = \tint \dd{x} |\, \Psi(x) \,|^2$,
we obtain
\begin{equation}
   H = \alpha^2 \, H_1 - \alpha^k \, H_2 \>,
\end{equation}
where
\begin{eqnarray}\label{H1H2}
   H_1
   &=
   \frac{1}{2} \tint \dd{y} |\,\partial_y \Psi(y) \,|^2 > 0 \>,
   \\
   H_2
   &=
   \frac{g}{k+1}\tint \dd{y} [\, \Psi^{\dag}(y) \Psi(y) \,]^{k+1} > 0 \>,
\end{eqnarray}
for all $k$
as well as
\begin{subeqnarray}\label{staticHder}
   \frac{\partial H(\alpha)}{\partial \alpha}
   &=
   2 \alpha H_1 - k \, \alpha^{k-1} \, H_2 \>,
   \\
   \frac{\partial^2 H(\alpha)}{\partial \alpha^2}
   &=
   2 \, H_1 - k(k-1) \alpha^{k-2} \, H_2 \>.
\end{subeqnarray}
Setting the first (partial) derivative to zero at $\alpha = 1$ gives an 
equation consistent with the equations of motion:
\begin{equation}\label{staticEOM}
    k H_2 = 2 H_1 \>,
\end{equation}
whereas the second derivative evaluated at the minimum, and at $\alpha = 1$
reads
\begin{equation}\label{DerrickTheorem}
   \frac{\partial^2 H(\alpha)}{\partial \alpha^2} \Big |_{\alpha=1}
   =
   k (2 - k) \, H_2\>.
\end{equation}
Thus, we see that at $k=1$, the exact solutions for the free case are 
stable. Only when $k >2$ do the solutions become unstable to scale transformations. 
However, once one adds the external complex potential terms, the windows of 
stability need to be determined by the stability curve $p(v)$ or by simulations
of the NLSE equation.

It should be noted in passing that for $k=1$ and using Eq.~\ef{staticsol}, we
have that
\begin{subeqnarray}\label{Hstaticsols}
   H_1
   &=
   (M_1 + M_2) \, \beta^2/3 \>,
   \\
   H_2
   &=
   g \, (M_1 + M_2)^2   \beta / 6 \>,
\end{subeqnarray}
so that imposing \ef{staticEOM} for $k=1$ gives $g \, (M_1 + M_2) = 4 \,\beta$.  
This is satisfied by the exact solution. 

%
\section{\label{s:collective}Collective coordinates}

We consider in this work external potentials of the form:
\begin{subeqnarray}\label{A-V}
   V_0(x)
   &= 
   a_1 \cos{k_1x} \>,
   \label{A-V-a} \\
   V_1(x)
   &= 
   a_2 \cos{k_2 x} \>,
   \label{A-V-b} \\   
   W(x)
   &= 
   a_3 \cos{k_3 x} \>,
\end{subeqnarray}
which are (all real and) even functions of $x$. For $V_0(x)$ to be confining 
near $x=0$ we need $a_1 < 0$. We review here the method of CCs 
(see for example Ref.~\cite{1751-8121-50-48-485205}) applied to our case. The 
time-dependent variational approximation relies on introducing a finite set 
of time-dependent real parameters in a trial wave function that 
hopefully captures the time evolution of a perturbed solution. By doing this, one obtains 
a simplified set of ordinary differential equations (ODEs) for the CCs 
in place of solving the full PDE for the NLS equation. 
To this end, let us first set
\begin{eqnarray}\label{e:VT-1}
   \Psi(x,t)
   &\mapsto
   \Psi[\,x,Q(t)\,] 
   \\
   Q(t) 
   &= 
   \{\, Q^1(t),Q^2(t),\dots,Q^{2n}(t) \,\} \in \mathbb{R}^{2n} \>,
   \notag
\end{eqnarray}
where $Q(t)$ are the CCs. We note that the success of the method depends 
greatly on the choice of the the trial wave function $\tilde{\Psi}[\,x,Q(t)\,]$. 
The generalized dissipative Euler-Lagrange equations lead to Hamilton's 
equations for $Q(t)$. 
The Lagrangian in terms of $Q(t)$ is given by 
\begin{equation}\label{e:VT-2}
   L(Q,\dot{Q})
   =
   T(Q,\dot{Q}) - H(Q) \>
\end{equation}
with the dynamic term 
\begin{eqnarray}
   T(Q,\dot{Q})
   &=
   \frac{\rmi}{2} \tint \dd{x}
   \bigl \{ \, 
      \Psi^{\dag}(x,Q)\, \Psi_t(x,Q)
      - 
      \Psi^{\dag}_t(x,Q) \, \Psi(x,Q) \,
   \bigr \}
   \notag \\
   &=
   \pi_\mu(Q) \, \dot{Q}^\mu \>,
   \label{Tdef}
\end{eqnarray}
and $\pi_{\mu(Q)}$ defined via
\begin{equation}\label{e:VT-3}
   \pi_\mu(Q)
   =
   \frac{\rmi}{2} \tint \dd{x}
   \bigl \{ \, 
      \Psi^{\dag}(x,Q)\,[\, \partial_\mu \Psi(x,Q) \,]
      - 
      [\, \partial_\mu \Psi^{\dag}(x,Q) \,] \, \Psi(x,Q) \,
   \bigr \} \>,
\end{equation}
where $\partial_{\mu} := \partial / \partial Q^{\mu}$.
The Hamiltonian $H(Q)$ is given by
\begin{equation}\label{e:VT-4}
   \fl
   H(Q)
   =
   \tint \dd{x} 
   \Bigl \{ \,
       |\partial_x \Psi(x,Q) |^2
       -
       \Psi^{\dag}(x,Q) \, U_0(x) \, \Psi(x,Q)
       -
       \frac{g}{2} \, |\Psi(x,Q)|^{4} \,
    \Bigr \} \>,
\end{equation}
and 
on an equal footing, the dissipation functional (again, in terms of the CCs) 
is respectively given by
\begin{eqnarray}
   \fl
   F(Q,\dot{Q})
   &=
   \rmi \tint \dd{x}  \,
   \bigl \{\,
      \Psi^{\dag}(x,Q) \, U_1(x) \, \Psi_t(x,Q)
      -
      \Psi_t^{\dag}(x,Q) \, U_1(x) \Psi(x,Q) \,
   \bigr \}
   \label{e:VT-4.1} \\ 
   \fl
   &=   
   w_{\mu}(Q) \, \dot{Q}^{\mu} \>,
   \notag
\end{eqnarray}
where
\begin{equation}\label{e:VT-4.2}
   \fl
   w_{\mu}(Q) 
   =   
   \rmi \tint \dd{x}  \,
   \bigl \{\,
      \Psi^{\dag}(x,Q) \, U_1(x) \, [\, \partial_{\mu} \Psi(x,Q) \,]
      -
      [\,  \partial_{\mu} \Psi_t^{\dag}(x,Q) \, \, U_1(x) \Psi(x,Q) \,
   \bigr \}
\end{equation}
with $U_0(x)$ and $U_1(x)$ being given by Eq.~\ef{U0U1def}.

This way, the generalized Euler-Lagrange equations read
\begin{equation}\label{e:VT-5}
   \frac{\partial L}{\partial Q^\mu}
   -
   \frac{\rmd}{\dd{t}} \Bigl ( \pdv{L}{\dot{Q}^\mu} \Bigr )
   =
   -
   \pdv{F}{\dot{Q}^\mu} \>.
\end{equation}
If $v_{\mu}(Q) :=\partial_\mu H(Q)$, we find 
\begin{equation}\label{e:VT-6}
   f_{\mu\nu}(Q) \, \dot{Q}^\nu
   =
   u_{\mu}(Q)
   =
   v_{\mu}(Q) - w_{\mu}(Q) \,,
\end{equation}
where 
\begin{equation}\label{e:VT-7}
   f_{\mu\nu}(Q)
   =
   \partial_\mu \pi_\nu(Q) - \partial_\nu \pi_\mu(Q)
\end{equation}
is an antisymmetric $2n \times 2n$ symplectic matrix.  
If $\det{(f(Q))} \ne 0$, we can define an inverse as the contra-variant matrix with upper indices,
\begin{equation}\label{e:VT-8}
   f^{\mu\nu}(Q) \, f_{\nu\sigma}(Q) = \delta^\mu_\sigma \>,
\end{equation}
in which case the equations of motion \ef{e:VT-6} can be put in the symplectic form:
\begin{equation}\label{e:VT-9}
   \dot{Q}^\mu
   =
   f^{\mu\nu}(Q) \, u_{\nu}(Q) \>.
\end{equation}

%
\section{\label{s:8CC}Eight parameter time-dependent collective coordinates}

From Eqs.~\ef{AntiPTeq} and \ef{Udef}, the coupled equations we wish to solve 
are given by
\begin{subeqnarray}\label{e:DEQ}
   \fl
   \{\,
      \rmi \,
      \partial_t
      +  
      \partial_x^2
      -
      V(x) \>\>
      + 
      g \, ( | \psi_1(x,t) |^{2}+ | \psi_2(x,t) |^{2}) \,
   \} \, \psi_1(x,t) 
   -
   \rmi\, W(x) \, \psi_2(x,t)
   &=
   0 \>,
   \label{DEQ-a} \\
   \fl
   \{\,
      \rmi \,
      \partial_t
      +  
      \partial_x^2
      -
      V^{\ast}(x)
      +
      g \, ( | \psi_1(x,t) |^{2}+ | \psi_2(x,t) |^{2}) \,
   \} \, \psi_2(x,t) 
   -
   \rmi\, W(x) \, \psi_1(x,t)
   &= 0 \>. 
   \label{DEQ-b}
\end{subeqnarray}
We choose time-dependent variational wave functions of the form:
\begin{subeqnarray}\label{t-psi12}
   \psi_1[x,Q(t)]
   &=
   A_1(t) \, \beta(t) \sech[\, \beta(t) (x - q(t)) \,] \, 
   \rme^{\rmi\, [\, \phi[x,Q(t)] - \theta_1(t) \,]} \>,
   \label{t-psi1} \\
   \psi_2[x,Q(t)]
   &=
   A_2(t) \, \beta(t) \sech[\, \beta(t) (x - q(t)) \,] \, 
   \rme^{\rmi\, [\, \phi[x,Q(t)] - \theta_2(t) \,]} \>,   
\end{subeqnarray}
where  
\begin{equation}\label{phidef}
   \phi[x,Q(t)]
   =
   p(t) \, (x - q(t)) + \Lambda(t) \, (x - q(t))^2 \>.
\end{equation}
For the variational solutions, we define
\begin{subeqnarray}\label{M12defs}
   M_1(t)
   &=:
   \int \dd{x} | \psi_1[x,Q(t)] |^2
   =
   2 \beta(t) \, |A_1(t)|^2 \>,
   \label{M1def} \\
   M_2(t)
   &=:
   \int \dd{x} | \psi_2[x,Q(t)] |^2
   =
   2 \beta(t) \, |A_2(t)|^2 \>.
   \label{M2def}   
\end{subeqnarray}
We will choose as our CCs 
the set of eight quantities:
\begin{equation}\label{AllQs}
   Q
   =
   \{\, M_1, \theta_1, M_2, \theta_2, q, p, \beta, \Lambda \, \} \>,
\end{equation}
with the canonical pairs,
\begin{equation}\label{CanonicalPairs}
   \{ M_1(t), \theta_1(t) \},
   \quad
   \{ M_2(t), \theta_2(t) \},
   \quad
   \{ q(t),p(t) \},
   \quad
   \{ \beta(t), \Lambda(t) \}.
\end{equation}
The CCs $Q(t)$ are related to the low order moments of the coordinate and
momentum operators so that their actual behavior can be determined from the
numerical simulation of the NLSE. This choice of CCs was determined after 
the numerical simulations suggested that the widths, position, and momenta 
of the two components followed one another closely (even though they were 
not exactly equal as we will see in our numerical simulations). 

%
\subsection{\label{ss:ICs}Initial conditions}

\begin{figure}[t]
\centering
\includegraphics[width=0.45\columnwidth]{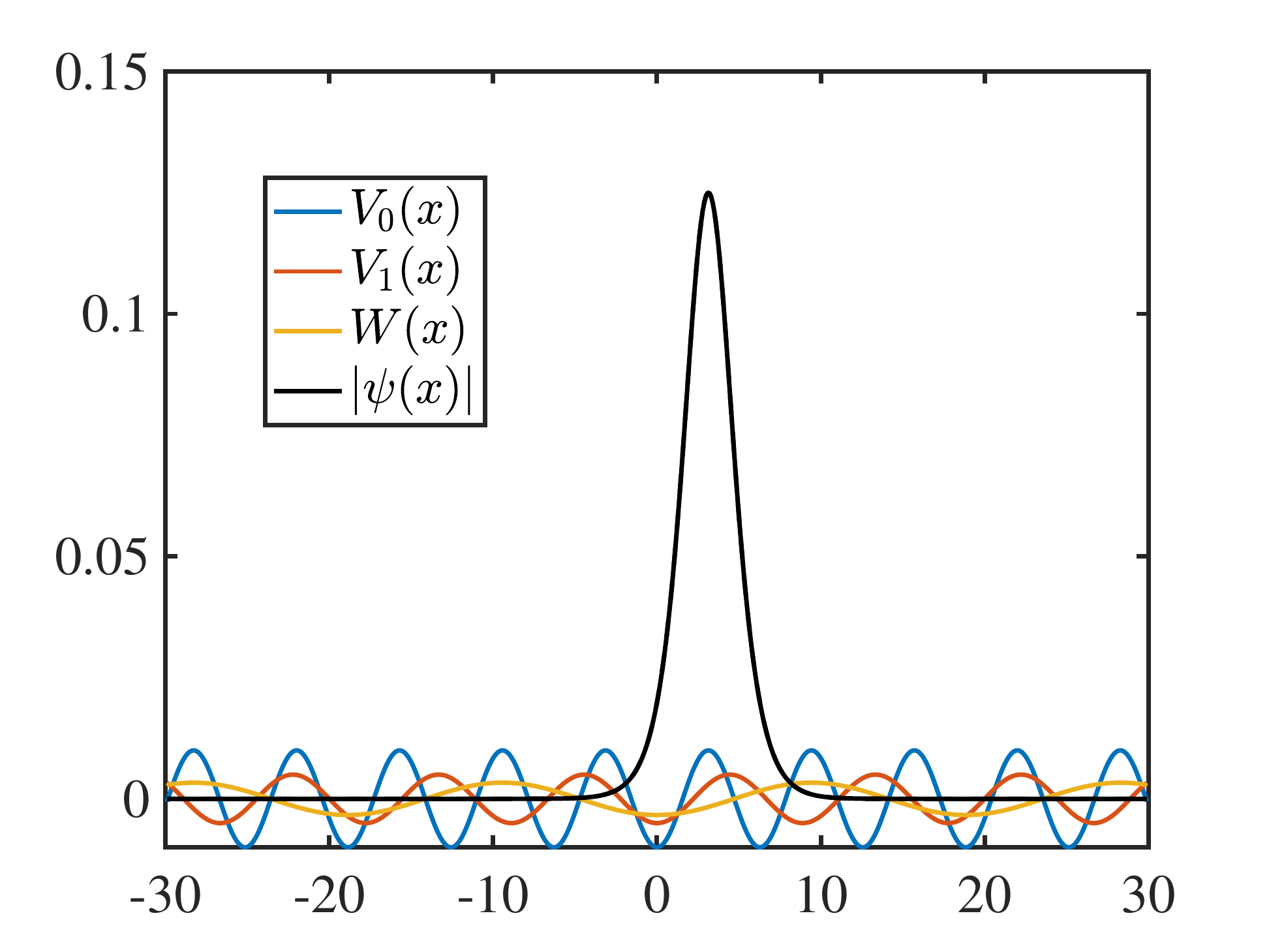}
\caption{\label{f:potentials} (Color online) Potentials and initial wave functions
for 8CC variational calculations for the parameters of Section~\ref{ss:ICs}
with $q(0) = \pi$.  Here we have set $a_1 = -1/100$, $a_2 = -1/200$, and 
$a_3 = - 1/300$ with $k_1 = 1$, $k_2 = 1/\sqrt{2}$, and $k_3 = 1/3$.}
\end{figure}

At $t=0$, we require that the variational wave functions [cf. Eqs.~\ef{t-psi12}]
match the traveling wave solution of Eq.~\ef{A-Psi} with no external potential. 
In addition, we require that initially the wave function is an eigenstate of the
anti-$\PT$ operator. Furthermore, we choose $g=2$ and $\beta(0)=1/2$ in order to
draw direct comparisons with our previous work on the NLSE in a $\PT$-symmetric 
potential~\cite{PhysRevE.94.032213}.
This means that at $t=0$ we set
\begin{eqnarray}\label{ICs1}
   \fl
   \beta(0) = 1/2
   \qc
   \Lambda(0) = 0
   \qc
   \theta_1(0) = 0
   \qc
   \theta_2(0) = \kappa \, \pi/2
   \qc
   M_1(0) = M_2(0) = 1/2,
\end{eqnarray}
so that $A_1(0) = A_2(0) = 1/\sqrt{2}$. Plots of the potentials and initial 
variational wave functions are shown in Fig.~\ref{f:potentials} where we have
set $q(0) = \pi$. Note that the magnitudes of the two wave functions are identical at $t=0$. 

%
%
%
%

%
\subsection{\label{ss:EoMI}Equations of motion}

Following the method described in Section~\ref{s:collective}, and using the 
variational wave function \ef{t-psi12}, we find the following equations of 
motion for the 8CCs: 
\begin{subeqnarray}\label{EOMX}
   \fl
   \dot{M}_1
   &=
   a_2 \, M_1 \, \cos(k_2 q) \, G_1(k_2/\beta)
   \label{EOMX-1} \\
   \fl
   & \hspace{1em}
   +
   a_3 \sqrt{M_1 M_2} \, \cos( k_3 q) \, \cos(\theta_1 - \theta_2) \,
   G_1(k_3/\beta) \>,
   \notag \\
   \fl
   \dot{\theta}_1
   &=
   -
   p^2
   +
   \frac{2}{3} \,\beta^2
   +
   \frac{a_1}{2} \, \cos(k_1 q) \, 
   \Bigl [\,
      G_1(k_1/\beta)
      -
      \frac{k_1}{2 \, \beta} \, G'_1(k_1/\beta) \,
   \Bigr ] 
   -
   \frac{5}{12} \, g \, \beta \, M 
   \notag \\
   \fl
   & \hspace{1em}
   +
   a_2 \, \frac{p}{\beta} \frac{M_1 - M_2}{M} \sin(k_2 q) \, G_3(k_2/\beta) \>
   \label{EOMX-2} \\
   \fl
   & \hspace{1em}
   -
   (a_3/2) \sqrt{M_2 / M_1} \, 
   \cos(k_3 q) \sin(\theta_1 - \theta_2) \, G_1(k_3/\beta)
   \notag \\
   \fl
   & \hspace{1em}
   +
   a_3 \sqrt{M_1 M_2 / M^2} \, (2p/\beta) \, 
   \sin(k_3 q) \cos(\theta_1 - \theta_2) \, G_3(k_3/\beta) \>,
   \notag \\
   \fl
   \dot{M}_2
   &=
   - a_2 \, M_2 \, \cos(k_2 q) \, G_1(k_2/\beta) \>
   \label{EOMX-3} \\
   \fl
   & \hspace{2em}
   +
   a_3 \sqrt{M_1 M_2} \, \cos( k_3 q) \, \cos(\theta_1 - \theta_2) \,
   G_1(k_3/\beta) \>,   
   \notag \\
   \fl
   \dot{\theta}_2
   &=
   -
   p^2
   +
   \frac{2}{3} \,\beta^2
   +
   \frac{a_1}{2} \, \cos(k_1 q) \, 
   \Bigl [\,
      G_1(k_1/\beta)
      -
      \frac{k_1}{2 \, \beta} \, G'_1(k_1/\beta) \,
   \Bigr ] 
   -
   \frac{5}{12} \, g \, \beta \, M 
   \notag \\
   \fl
   & \hspace{1em}
   +
   a_2 \, \frac{p}{\beta} \frac{M_1 - M_2}{M} \sin(k_2 q) \, G_3(k_2/\beta) \>
   \label{EOMX-4} \\
   \fl
   & \hspace{1em}
   -
   (a_3/2) \sqrt{M_1 / M_2} \, 
   \cos(k_3 q) \sin(\theta_1 - \theta_2) \, G_1(k_3/\beta)
   \notag \\
   \fl
   & \hspace{1em}
   +
   a_3 \sqrt{M_1 M_2 / M^2} \, (2p/\beta) \, 
   \sin(k_3 q) \cos(\theta_1 - \theta_2) \, G_3(k_3/\beta) \>,
   \notag \\
   \fl
   \dot{q}
   &=
   2 \, p
   -
   \frac{a_2}{\beta} \, \frac{M_1 - M_2}{M} \sin(k_2 q) \, G_3(k_2/\beta)
   \label{EOMX-5} \\
   \fl
   & \hspace{2em}
   -
   a_3 \sqrt{M_1 M_2/M^2} \, \sin(k_3 q) \, 
   \cos(\theta_1 - \theta_2) \, (2/\beta) \, G_3(k_3/\beta) \>,
   \notag \\
   \fl
   \dot{p}
   &=
   \frac{k_1 a_1}{2} \, \sin(k_1 q) \, G_1(k_1/\beta)
   -
   a_2 \, \frac{M_1 - M_2}{M} \,
   \frac{2 \Lambda}{\beta} \sin(k_2 q) \, G_3(k_2/\beta)
   \label{EOMX-6} \\
   \fl
   & \hspace{2em}
   -
   a_3 (4 \Lambda/\beta) \, \sqrt{M_1 M_2/M^2} \,
   \sin(k_3 q) \cos(\theta_1 - \theta_2) \, G_3(k_3/\beta) \>,
   \notag \\
   \fl
   \dot{\beta}
   &=
   - 
   4 \, \beta \, \Lambda
   +
   \frac{M_1 - M_2}{M} \, a_2 \cos(k_2 q) \, \frac{\beta}{2} \, 
   \bigl [\, 
      G_1(k_2/\beta)
      -
      (12/\pi^2) \, G_2(k_2/\beta) \,
   \bigr ] \>
   \label{EOMX-7} \\
   \fl
   & \hspace{0em}
   +
   a_3 \beta \sqrt{M_1 M_2/M^2} \,
   \cos(k_3 q) \cos(\theta_1 - \theta_2) \,
   \bigl [\,
      G_1(k_3/\beta) - (12/\pi^2) \, G_2(k_3/\beta) \,
   \bigr ] \>,
   \notag \\
   \fl
   \dot{\Lambda}
   &=
   -
   4 \, \Lambda^2
  +  \frac{4 \, \beta^4}{\pi^2}
   +
   a_1 \, \frac{6 \, k_1 \, \beta}{\pi^2} \, G'_1(k_1/\beta)
   -
   \frac{g \, \beta^3}{\pi^2} \, M \>.    
   \label{EOMX-8}
\end{subeqnarray}
Details of this derivation are given in \ref{s:8CCderivation}.

%
\subsection{General Observations about the 8CC equations} 

Firstly, we note that $M_1$ and $M_2$ go in opposite directions from 
their initial yet equal value due to $a_2 \neq 0$. This often leads to one 
of the two masses going to zero. 
We further note that when $a_2=0$, the effect of $a_3$ on the dynamics 
is proportional to $\cos (\theta_1-\theta_2)$ which initially is zero. 
Moreover, $M_1(0) = M_2(0)$ due to anti-$\PT$ initial conditions.
The equation for $\dot{\theta}_1 - \dot{\theta}_2$ is given by 
\begin{equation}
   \fl
   \dot{\theta}_1 - \dot{\theta}_2   
   = 
   \frac{\pi a_3} { 2 \beta} \,
   \Biggl [ 
      \Biggl ( 
         \sqrt{ \frac{M_2}{M_1}}
         -
         \sqrt{\frac{M_1 }{M_2}}  
      \Biggr ) 
      \cos( q(t)/4 ) \csch (\pi/(8 \beta)) 
      \sin (\theta_1-\theta_2) 
   \Biggr ] \>.
\end{equation}
Since the derivative is initially zero because the two masses are the same
(unless $M_1$ differs greatly from $M_2$), $ \theta_1 -\theta_2$ stays small, 
and the presence of $a_3$ does not change the CC equations for $q, p,\beta ,M_1,M_2$ 
greatly from the case when  $a_3 = 0$. 

%
\subsection{\label{ss:smallosc}Small oscillation equations when $a_2=a_3=0$}

When $a_2 <0 $, 
$M_{1}$ and $M_{2}$ initially decrease and increase with time, respectively 
(or vice versa depending on the sign of $\cos k_2 x_0$),
so one is never in the small oscillation regime. However when $a_2 = a_3 = 0$ small
oscillations are possible in the potential $V_0(x) $.  In the small deviation 
from the static soliton regime, the update equations for the set $(q,p)$ decouple 
from the set $(\beta, \Lambda)$.  The relevant equations when $a_2=a_3=0$ are
\begin{eqnarray}\label{SO-1}
   \dot{q} 
   &= 2\, p
   \qc
   \dot{p} 
   = 
   \frac{k_1 a_1}{2} \sin(k_1 q) \, G_1(k_1/\beta) \,, 
   \\
   \dot{\beta}
   &=
   -  4 \, \beta \, \Lambda
   \qc
   \dot{\Lambda}
   =
   -
   4 \, \Lambda^2
   +
   \frac{4 \, \beta^4}{\pi^2}
   +
   a_1 \, \frac{6 \, k_1 \, \beta}{\pi^2} \, G'_1(k_1/\beta)
   -
   \frac{g \, \beta^3}{\pi^2} \, M \>.
   \notag
\end{eqnarray}
Setting $\beta(t) = 1/2 + \delta \beta(t)$ with $\delta \beta(t)\ll1\, \forall t$ 
(and all the other parameters assumed small deviations from zero), 
one has for the first two equations in \ef{SO-1}:
\begin{equation}\label{SO-2}
   \dot{q}(t) = 2 \, p(t)
   \qc
   \dot{p}(t) = [\, a_1 k_1^3 \pi \csch(k_1 \pi) \, ] \,  q(t) \>.
\end{equation}
Since $a_1 <0$, we have that the frequency of both $p$ and $q$ (in
this small oscillation regime) is just
\begin{equation}
   \omega^2_q = 2 |a_1| k_1^3 \pi  \csch (k_1 \pi) \>.
\end{equation}
For instance, if $a_1 = -1/100$ and $k_1=1$, the period $T_{q}$ 
is given by
\begin{equation}
   T_q
   = 
   \frac{2 \,\pi}{\omega_{q}}
   =
   \sqrt{ \frac{2 \pi}{a_1  k_1^3 \csch(\pi  k_1)} }
   \approx 
   85.2 \>.
\end{equation}
Since initially  $g M= 4 \,\beta(0)= 2$, we find (ignoring the $a_1$ correction) 
\begin{equation}
   \delta\dot{\beta} = - 2 \, \delta \Lambda
   \qc
   \delta\dot{\Lambda} = \frac{\delta \beta }{2 \,\pi^2},
\end{equation}
such that 
\begin{equation}
   \omega_{\beta} = \frac{1}{\pi}
   \qc
   T_{\beta} = \frac{2 \pi}{\omega_{\beta}} = 2 \pi^2  \approx 19.7392 \>.
\end{equation}
To include the $a_1$ correction, one can use
\begin{eqnarray}
   \fl
   G'_1(k_1/\beta)
   &\rightarrow 
   2 \pi ^2  k_1 \,\delta \beta \,
   [\, 
      \pi k_1-2 \pi k_1 \coth ^2(\pi k_1) 
      + 
      2 \coth (\pi k_1) \,
   ] \, \csch(\pi k_1)
   \notag \\
   \fl
   &\hspace{5em}
   + \pi \, [\, 1 - \pi k_1 \coth (\pi k_1) \,] \csch(\pi k_1).
\end{eqnarray}

%
\section{\label{s:comp}Comparison of Numerical Simulations with 8CC equations evolution}
In solving for the time evolution of the NLSE in these external potentials, 
we will employ initial conditions corresponding to the exact solution of the 
NLSE in the free case. 
The 
configuration space of possible solutions (and their associated time evolution) 
is huge, and we will just exhibit five cases to give the general idea of how well 
the CC approach matches with the time evolution of the NLSE.  We have chosen 
parameters to be similar to those used in our previous 
work on the single-component $\PT$-symmetric NLSE. 

The cases we 
study hereafter are presented in Table~\ref{t:parameters} and summarize 
several behaviors we identified.
We have chosen $q_0$ so that as far as $V_0$ is concerned, the initial wave function is starting at either a minimum of the 
potential ($q_0=0$), or a maximum of the potential ($q_0=\pi$).
In particular, in cases 1  and 5, the soliton is trapped by the 
potential $V_0$. In case 1, all $k_i$ are  different and $q_0= \pi$. In 
case 5 we have instead $k_i=1,~ q_0=0$.  Case 2 is a moving soliton that 
is unstable. Case 3 shows the effect of $a_3$ on a moving soliton when $a_2=0$. 
To first-order approximation the result is similar to the case where $a_3=0$ 
in that the width of both components just oscillates, and (at least for a reasonable
amount of time) the two components stay equal in mass and these masses do not 
change in time.  
Case 4 shows what happens when we add $a_2$ to case 3, 
which then causes $M_1$ to gradually increase, and $M_2$ to gradually decrease. 
This situation is unstable as the total mass $M_1+M_2$ gradually increases. The 
initial values of the  parameters we use for the CC simulations are also given in 
Table~\ref{t:parameters}. These parameters also determine the initial wave function 
used in our numerical simulations. The values of $q_0$ and $p_0$ were chosen so 
that a comparison with simulations in the one-component case could be made. If 
we increase $a_2$ in magnitude much beyond $|a_2| = 1/300$, then the instabilities 
manifest themselves at quite earlier times. 

%
%
\begin{table}
\caption{\label{t:parameters}Parameters for simulations.  In all cases
we take $g=2$, $M_1(0) = M_2(0) = 1/2$, $\beta_1(0) = \beta_2(0) = 1/2$, 
and $\Lambda_1(0) = \Lambda_2(0) = 0$, and with $\theta_1(0) = 0$, $\theta_2(0) = \pi/2$.}
\begin{indented}
\item[]
\begin{tabular}{@{}lllllllll} 
\br
Case&$a_1$&$a_2$&$a_3$&$k_1$&$k_2$&$k_3$&$q(0)$&$p(0)$\\
\mr
1 & $-1/100$ & $-1/500$ & $-1/1000$ & $1$ & $1/\sqrt{2}$ & $1/3$ & $\pi$ & $.001$ \\
2 & $-1/100$ & $-1/100$ & $-1/500$ & $1$ & $1/3$ & $1/4$ & $\pi$ & $-0.0457$ \\
3 & $-1/100$ & $\phantom{-}0$ & $-1/100$ & $1$ & $1$ & $1$ & $\pi$ & $0.0531649$ \\
4 & $-1/100$ & $-1/1000$ & $-1/100$ & $1$ & $1$ & $1$ & $\pi$ & $0.0531649$ \\
5 & $-1/100$ & $-1/1000$ & $-1/100$ & $1$ & $1$ & $1$ & $0$ & $ 0.0531649$ \\
\br
\end{tabular}
\end{indented}
\end{table}
%
%

The cases shown in Table~\ref{t:parameters} are explored by performing 
numerical simulations at the level of Eqs.~(\ref{DEQ-a})-(\ref{DEQ-b}). 
At first, the infinite spatial domain is truncated into a finite one 
$[-L,L]$, and then a one-dimensional spatial grid of equidistant points 
with resolution $\Delta x$ is introduced ($L=30$ and $\Delta x=0.1$ in this work). 
The Laplacian in Eqs.~(\ref{DEQ-a})-(\ref{DEQ-b}) 
is replaced by a second-order accurate, finite difference scheme. We impose
zero Dirichlet boundary conditions at the edges of our computational domain, 
that is, $\psi_{1,2}(x=\pm L,t)=0$, $\forall t\geq 0$. As a result, the 
coupled NLSEs reduce into a (large) system of ODEs that are advanced forward
in time by employing the Dormand and Prince method with time step-size adaptation~\cite{Hairer}. 
When the dynamics revealed an instability of the pertinent waveforms, we 
stopped the integrator before they hit the boundary. Also, we corroborated our
numerical results by considering a fourth-order accurate, finite difference 
scheme for the Laplacian operator. We found that both discretization schemes 
produce identical results. 
%
%

To compare the numerical simulation results of the NLSEs with the 8CC equations 
we use the fact that we can extract the values of the CCs 
from
the various low order moments of the numerically obtained wave function. In fact,
the equations the low order moment equations obey are an alternative way of
obtaining equations that are equivalent to those obtained from the variational 
approach. Assuming a more general variational wave function \ansatz\ where we allow 
different values for the expectation of $x p, x^2, p, px$ for each component of
the wave function, we can extract easily 
the values of all these time evolving parameters from the moments of the numerical solution.  
In particular, 
let us assume that each component of the wave function can be parametrized as
\begin{eqnarray}
   \psi_i[x,Q(t)]
   &=
   A_i(t) \, \beta_i (t) \sech[\, \beta_i(t) (x - q_i(t)) \,] \, 
   \rme^{\rmi\, [\, \phi_i[x,Q(t)] - \theta_i(t) \,]} \>,
   \notag \\
   \phi_i[ x,Q(t)]
   &=
   p_i(t) \, (x - q_i(t)) + \Lambda_i(t) \, (x - q_i(t))^2 \>.
   \label{simpsi}
\end{eqnarray}
The $n^{\mathrm{th}}$ moment of the density distribution for each component 
of the wave function is defined by
\begin{eqnarray}\label{xmoments}
   \mathcal{M}^{i} _n(t)
   &=
   \tint \dd{x} x^n \, |\, \psi_{i}(x,t) \,|^2 
   \notag \\
   &=
   \frac{M_{i}(t)}{2}
   \tint \dd{y} \Bigl [\, \frac{y}{\beta_{i}(t)} + q_{i}(t) \, \Bigr ]^n \,
   \sech^2(y) \>,
\end{eqnarray}
which gives
\begin{subeqnarray}\label{nthxmoment}
   \mathcal{M}^{i}_0(t) &= M_{i}(t) \>,
   \label{xmoment-0} \\[5pt]
   \mathcal{M}^{i}_1(t) &= M_{i}(t) \, q_{i}(t) \>,
   \label{xmoment-1} \\
   \mathcal{M}^{i}_2(t) &= M_{i}(t) \,
   \Bigl [\, q_{i}^2(t) + \frac{\pi^2}{12} \, \frac{1}{\beta_{i} ^2(t)} \, \Bigr ] \>.
   \label{xmoment-2}
\end{subeqnarray}
Note that from Eqs.~\ef{nthxmoment}, we can find $M_i(t)$, $q_i(t)$, and $\beta_i(t)$.
On an equal footing, the $n^{\mathrm{th}}$ moment of the momentum operator is defined by
\begin{eqnarray}
   \mathcal{P}^{i}_n(t)
   &=
   \frac{1}{2 \rmi} \tint \dd{x} x^n
   \bigl \{\,
      \psi_i^{\ast}(x,t) \, [\, \partial_x \psi_i(x,t) \,]
      -
      [\, \partial_x \psi_i^{\ast}(x,t)\,] \, \psi_i(x,t) \,
   \bigr \}
   \notag \\
   &=
   \tint \dd{x} x^n
   \Imag{ \psi_i^{\ast}(x,t) \, [\, \partial_x \psi_i(x,t) \,] } \,, 
   \label{pmoments}
\end{eqnarray}
which gives
\begin{subeqnarray}\label{nthpmoment}
   \mathcal{P}^{i}_0(t) &= M_{i}(t) \, p_{i}(t) \>,
   \label{pmoment-0} \\[5pt]
   \mathcal{P}^{i}_1(t) &=
   M_{i}(t) \, 
   \Bigl [\, 
      p_{i}(t) \, q_{i}(t) 
      + 
      \frac{\pi^2 \Lambda_{i}(t)}{6 \, \beta_{i} ^2(t)} \,
   \Bigr ] \>,   
   \label{pmoment-1}
\end{subeqnarray}
from which we can find $p_i(t)$ and $\Lambda_i(t)$.
Finally, for the phase, we compute:
\begin{eqnarray}
   \mathcal{E}^{i}_0(t)
   &=
   \frac{\rmi}{2} \tint \dd{x}
   \bigl \{ \,
      \psi_i^{\ast}(x,t) \, [\, \partial_t \psi_i(x,t) \,]
      -
      [\, \partial_t \psi_i^{\ast}(x,t) \,] \, \psi_i(x,t) \,
   \bigr \}
   \notag\\
   &=
   M_i(t) \,
   \Bigl \{\, 
      \Bigl [ \,
         p_i(t) - \frac{\pi^2}{6} \, \frac{\Lambda_i(t)}{\beta_i(t)} \,
      \Bigr ] \, \dot{q}_i(t)
      + 
      \dot{\theta}_i(t) \,
   \Bigr \} \>, 
   \label{phase}
\end{eqnarray}
from which we can find $\dot{\theta_i}(t)$. 
We expect the time evolution of the higher moments of the coordinate and 
momentum operators (i.e. $\beta$ and $\Lambda$ of our variational 
ans\"atz) to become less accurate than the time evolution of the  lower moments, which seems 
to be the case in our simulations. 
What is remarkable is that to a good approximation, we find 
that using the moments of the numerical simulations of the wave function, the 
moments have the property that
\begin{equation}
   q_1(t) = q_2(t)
   \qc
   p_1(t) = p_2(t)
   \qc
   \beta_1(t) = \beta_2(t)
   \qc
   \Lambda_1(t) = \Lambda_2(t) \>,
\end{equation}
so one can use a trial wave function with 8 instead of 12 CCs. 

%
%
\begin{figure}
\centering
\includegraphics[width=0.95\columnwidth]{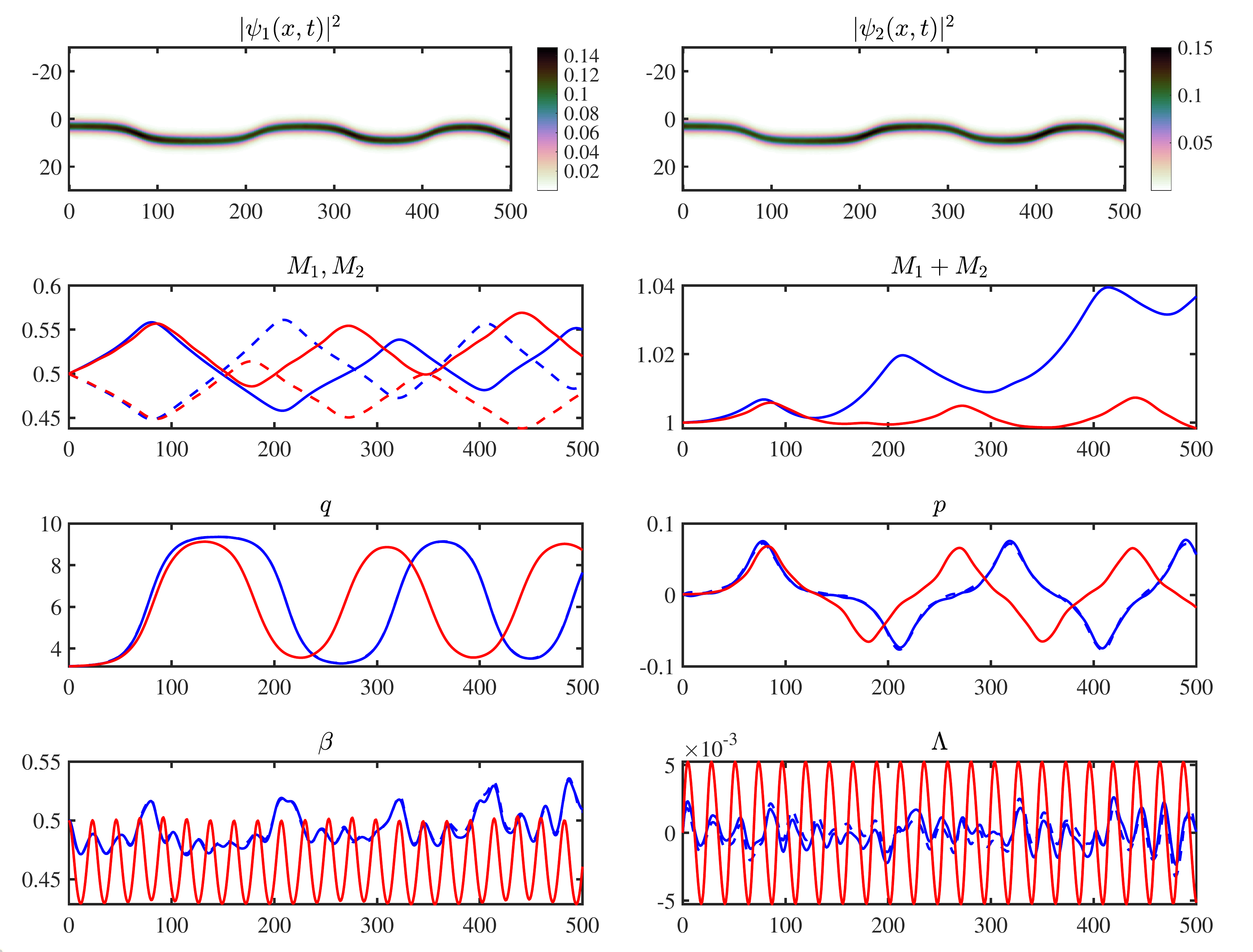}
\caption{\label{f:case1}
(Color online) 
Numerical results corresponding to parameters and initial conditions for case 1
of Table~\ref{t:parameters}. The top left and right panels demonstrate the 
spatio-temporal evolution of the densities $|\psi_{1}|^{2}$ and $|\psi_{2}|^{2}$,
respectively. The blue lines in the second, third, and fourth rows correspond 
to numerical results of the \Schrodinger's equation whereas the red lines to 
the 8CC variational calculation. 
The solid and dashed lines 
correspond to the first and second component, respectively. 
We see that around $t=200$ the variational approximation starts diverging quantitatively 
from the numerical result.} 
\end{figure}
%
%

%
%
\begin{figure}
\centering
\includegraphics[width=0.95\columnwidth]{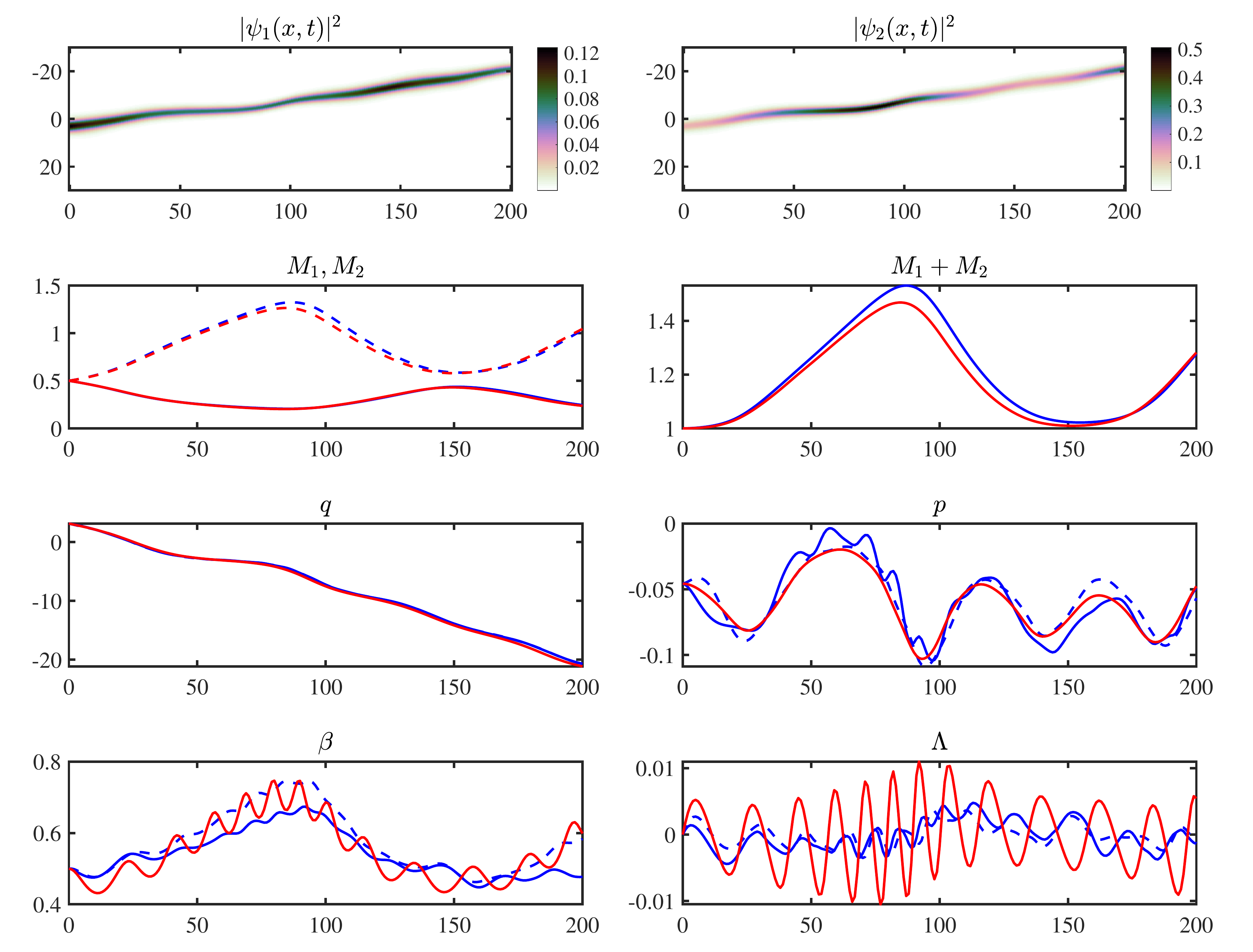}
\caption{\label{f:case2}
(Color online) 
Same as Fig.~\ref{f:case1} but for the case 2 of Table~\ref{t:parameters}.
The top left and right panels demonstrate the spatio-temporal evolution 
of the densities $|\psi_{1}|^{2}$ and $|\psi_{2}|^{2}$, respectively. 
The blue lines in the second, third, and fourth rows correspond to 
numerical results of the \Schrodinger's equation whereas the red lines 
to the 8CC variational calculation. The solid and dashed lines 
correspond to the first and second component, respectively.}
\end{figure}
%
%

%
%
\begin{figure}
\centering
\includegraphics[width=0.95\columnwidth]{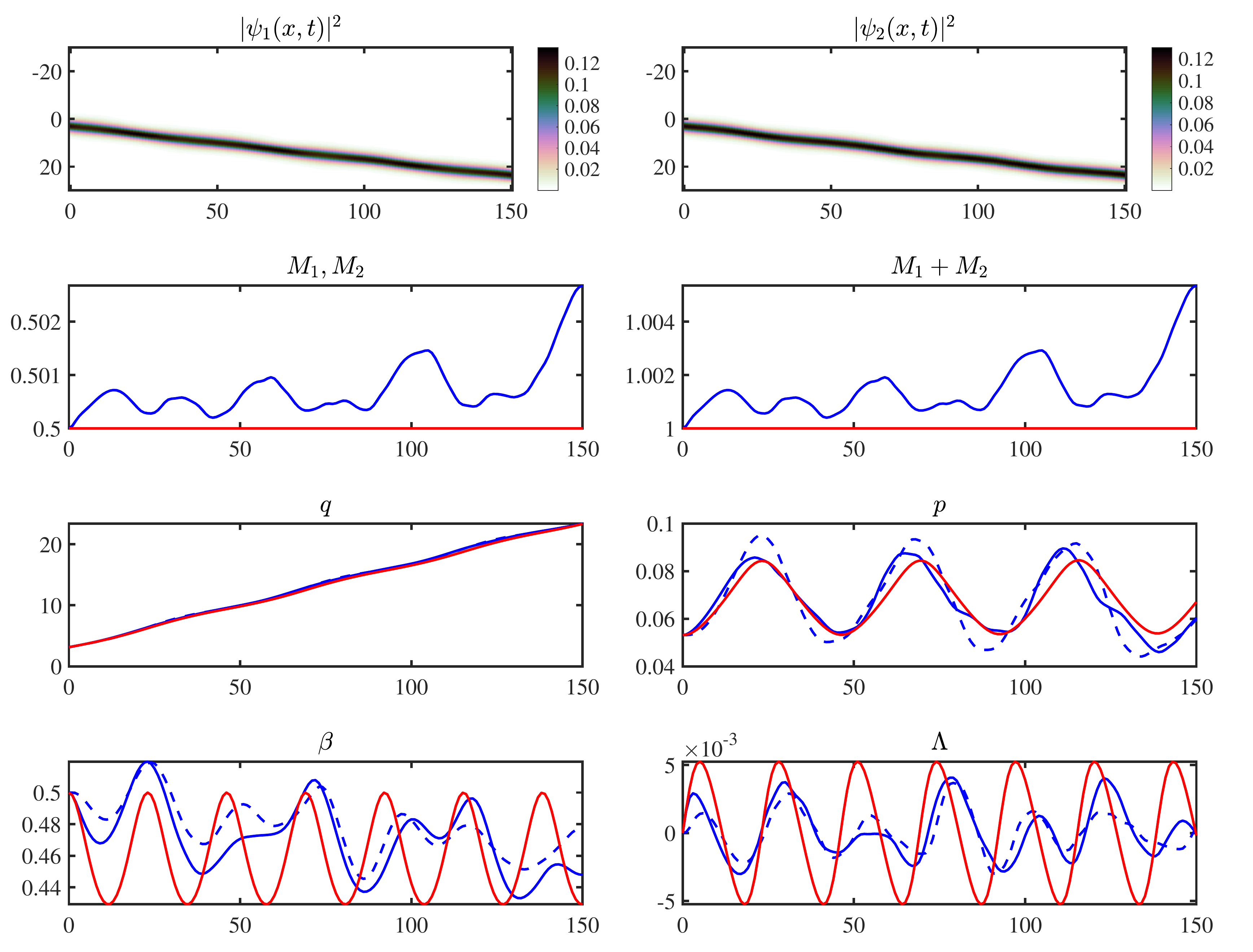}
\caption{\label{f:case3}
(Color online) 
Same as Fig.~\ref{f:case1} but for the case 3 of Table~\ref{t:parameters}.
Again, the top left and right panels demonstrate the spatio-temporal evolution 
of the densities $|\psi_{1}|^{2}$ and $|\psi_{2}|^{2}$, respectively. 
The blue lines in the second, third, and fourth rows correspond to 
numerical results of the \Schrodinger's equation whereas the red lines 
to the 8CC variational calculation. Finally, the solid and dashed lines 
correspond to the first and second component, respectively.}
\end{figure}
%
%

%
%
\begin{figure}
\centering
\includegraphics[width=0.95\columnwidth]{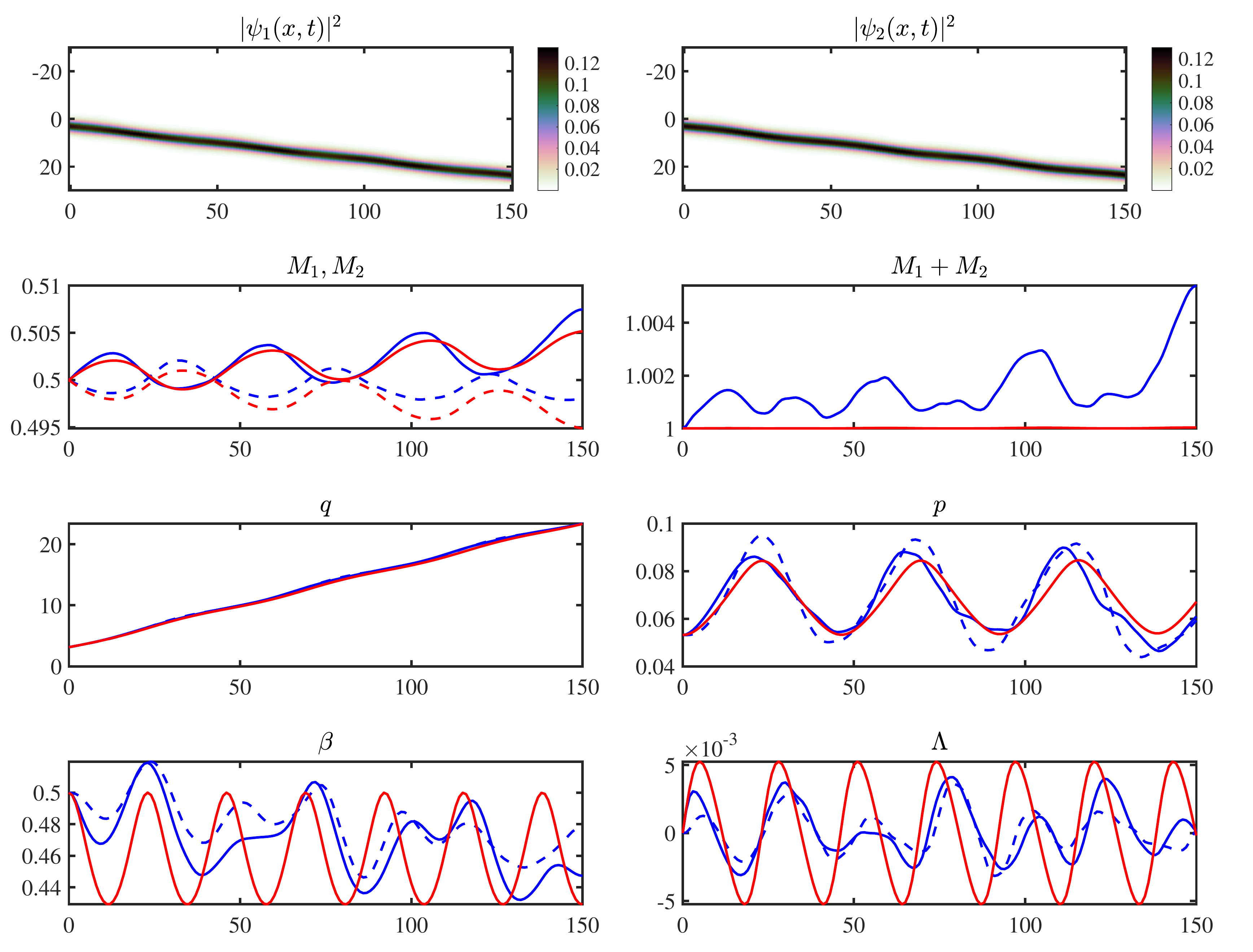}
\caption{\label{f:case4}
(Color online) 
Same as Fig.~\ref{f:case1} but for the case 4 of Table~\ref{t:parameters}.
The top left and right panels demonstrate the spatio-temporal evolution 
of the densities $|\psi_{1}|^{2}$ and $|\psi_{2}|^{2}$, respectively. 
The blue lines in the second, third, and fourth rows correspond to 
numerical results of the \Schrodinger's equation whereas the red lines 
to the 8CC variational calculation. Again, the solid and dashed lines 
correspond to the first and second component, respectively.
}
\end{figure}
%
%
\begin{figure}
\centering
\includegraphics[width=0.95\columnwidth]{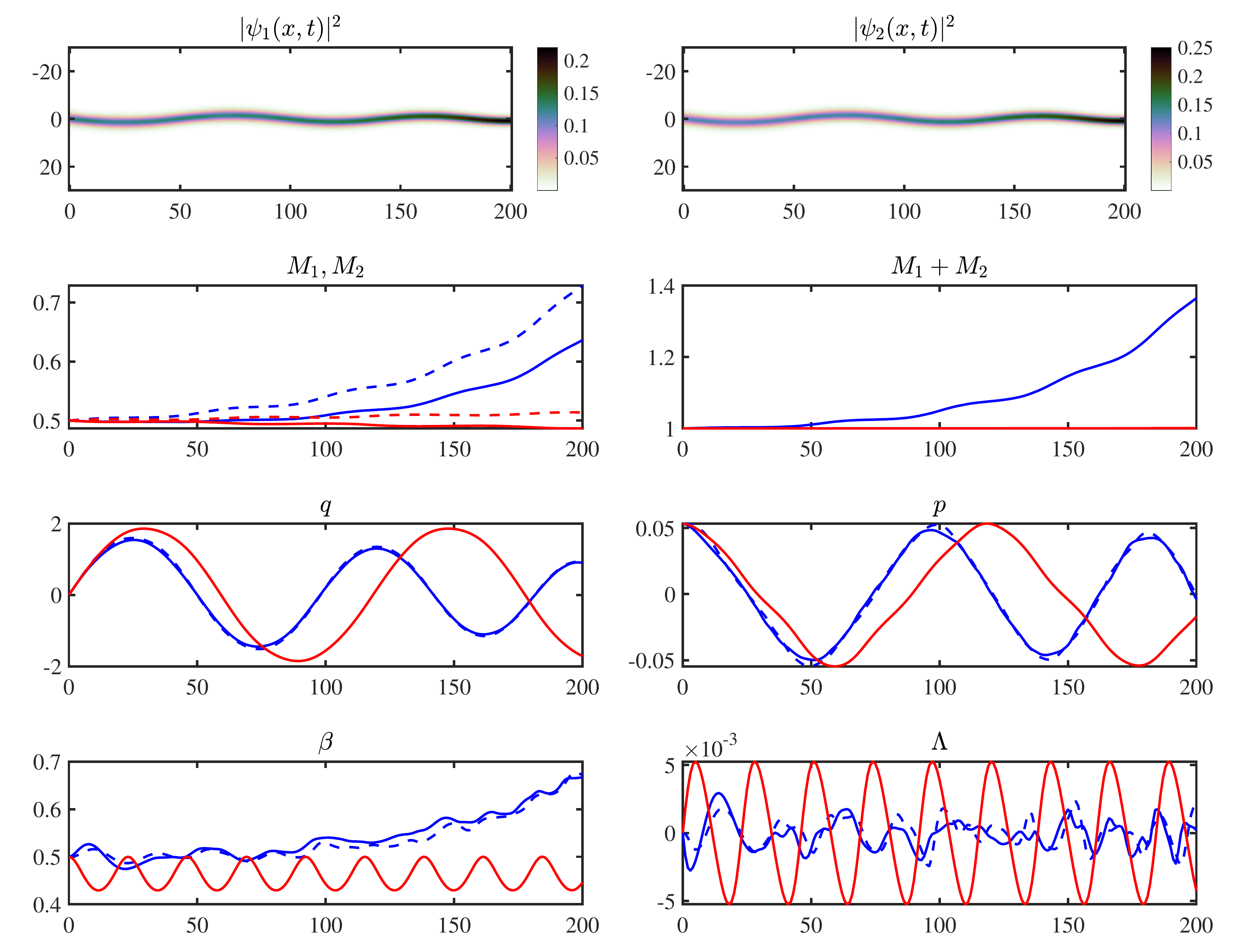}
\caption{\label{f:case5}
(Color online) 
Same as Fig.~\ref{f:case1} but for the case 5 of Table~\ref{t:parameters}.
All panels have the same format as the ones of Figs.~\ref{f:case2}-\ref{f:case4}.
}
\end{figure}

%
\section{\label{ss:typical} Discussion of Typical Behaviors}

In this section, we show some typical behaviors which are quite dependent 
on the parameters chosen (see Table~\ref{t:parameters}). 
If one looks at the potential $V_0(x)$ in Fig.~\ref{f:potentials}, we see 
it has maxima at $x= \pi $ and $x=3 \pi $ (in general at $x=(2n+1)\pi/k_{1}$ with $n\in\mathbb{Z}$) 
so if the soliton has a small initial momentum in the positive direction it can
lead to the behavior seen in 
Fig.~\ref{f:case1}. For this case, the soliton stays trapped between 
$\pi < x  <  3 \pi $ (see the panel showcasing $q(t)$ therein). 
At later times ($t> 300$) in the CC evolution one sees 
a very slight reduction in amplitude of the $q$ oscillations. Note that $\beta(t)$ 
continues to oscillate about $\beta(t) = 0.47$ and $\Lambda(t)$ about zero. 
Also, $M_1(t)$ is creeping up linearly with a very small slope, and $M_2(t)$ is 
decreasing linearly with a small slope such that the time averaged value 
of $M_1+M_2$ is remaining near one.  However the amplitudes of oscillations 
of $M_1+M_2$ have almost reached one percent by $t=100$. Here $p(v)$ indicates 
that this case is dynamically unstable as seen in Fig. \ref{pvplots}. When
we compare the CC results  to the numerical  simulations, we find that the
CCs are much closer to the numerical results for the lower order moments, 
but even $\beta(t)$ and $\Lambda(t)$ give qualitatively good results. We 
notice that $\beta$ and $\Lambda$ have a secondary oscillation frequency 
that is not captured by the CC equations. This is typical of what happens 
when the soliton is trapped by $V_0(x)$. 

The second example 
is shown in Fig.~\ref{f:case2} and corresponds to case 2 of Table~\ref{t:parameters}. 
Here, we chose different periods for the three potentials. This is a moving 
soliton where now $M_1(t)$ is decreasing slowly in time and $M_2(t)$ increasing in time. 
Here $\beta(t)$ as well as $M_1+M_2$ are increasing in time indicating eventual
blowup of the solitary wave.  The magnitudes of the oscillations of $p(t)$ 
and $\dot{q}(t)$ are decreasing in time, and at each turnaround $dp/dv <0$, 
thus indicating an unstable case. 
This behavior of $p(v)$ is shown in Fig.~\ref{pvplots}. 

Case 3 is shown in Fig.~\ref{f:case3}.  Here we consider the effect of $a_2$ 
on a moving soliton when $k_i =1$ for $i=1,2,3$. 
The 8CC approximation in this case gives 
$M_1 = M_2 = 1/2$ for all time so that the effect of $a_2$ on the motion in the 
real potential $V_0(x)$ is minimal. The actual numerics show that the 8CC approximation
is breaking down although 
the parameters $q(t)$, $p(t)$, $\beta(t)$, and $\Lambda(t)$ are qualitatively 
the same for both components  
(in fact, they differ so that the total mass $M_1+M_2$ very slowly increases).  

Case 4 is shown in Fig.~\ref{f:case4}.  Here we have a moving soliton starting 
at $q_0=\pi$ and the same initial conditions as in case 3 but we now turn on 
$a_2 = -1/1000$. This causes $M_1$ to slowly increase, and $M_2$ to slowly decrease, 
with $M_1+M_2$ slowly increasing which eventually leads to blowup. This instability 
is seen in the $p(v)$ curve shown in \ref{pvplots}. Here we start seeing a divergence 
from the solid blue lines for $q(t)$, $p(t)$, $\beta(t)$, and $\Lambda(t)$ from the
dashed blue lines in the numerical simulations, indicating a slight breakdown in 
our assumption that the two components have the same values. Nevertheless the 8CC 
parameters follow reasonably well the numerically obtained moments.

Case 5 is shown in Fig.~\ref{f:case5}.  Here we have a moving soliton starting at 
$q_0=0$ but otherwise the same initial conditions as case 4. This results in the 
soliton being trapped in the well of $V_0(x)$. Here $M_1$ slowly decreases and $M_2$ 
slowly increases, opposite to that in case 4, with $M_1 + M_2$ as well as $\beta(t)$ 
slowly increasing, which eventually leads to blowup. This instability is seen in the
$p(v)$ curve shown in \ref{pvplots}. 
  
%
\section{\label{s:pvcurve}Dynamical stability using the stability curve $p(v)$}

\begin{figure}
\centering
\includegraphics[width=0.65\columnwidth]{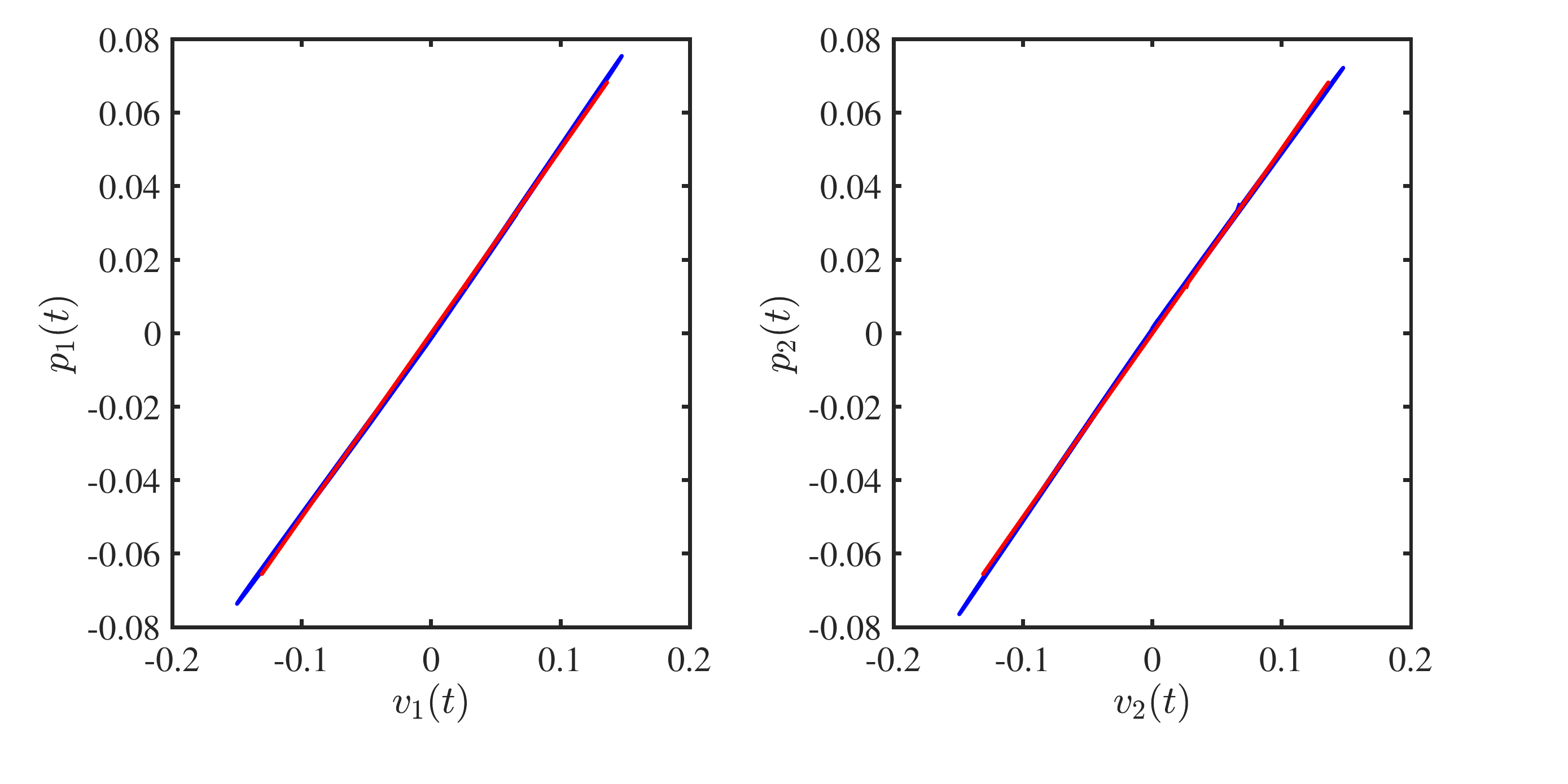}
\includegraphics[width=0.65\columnwidth]{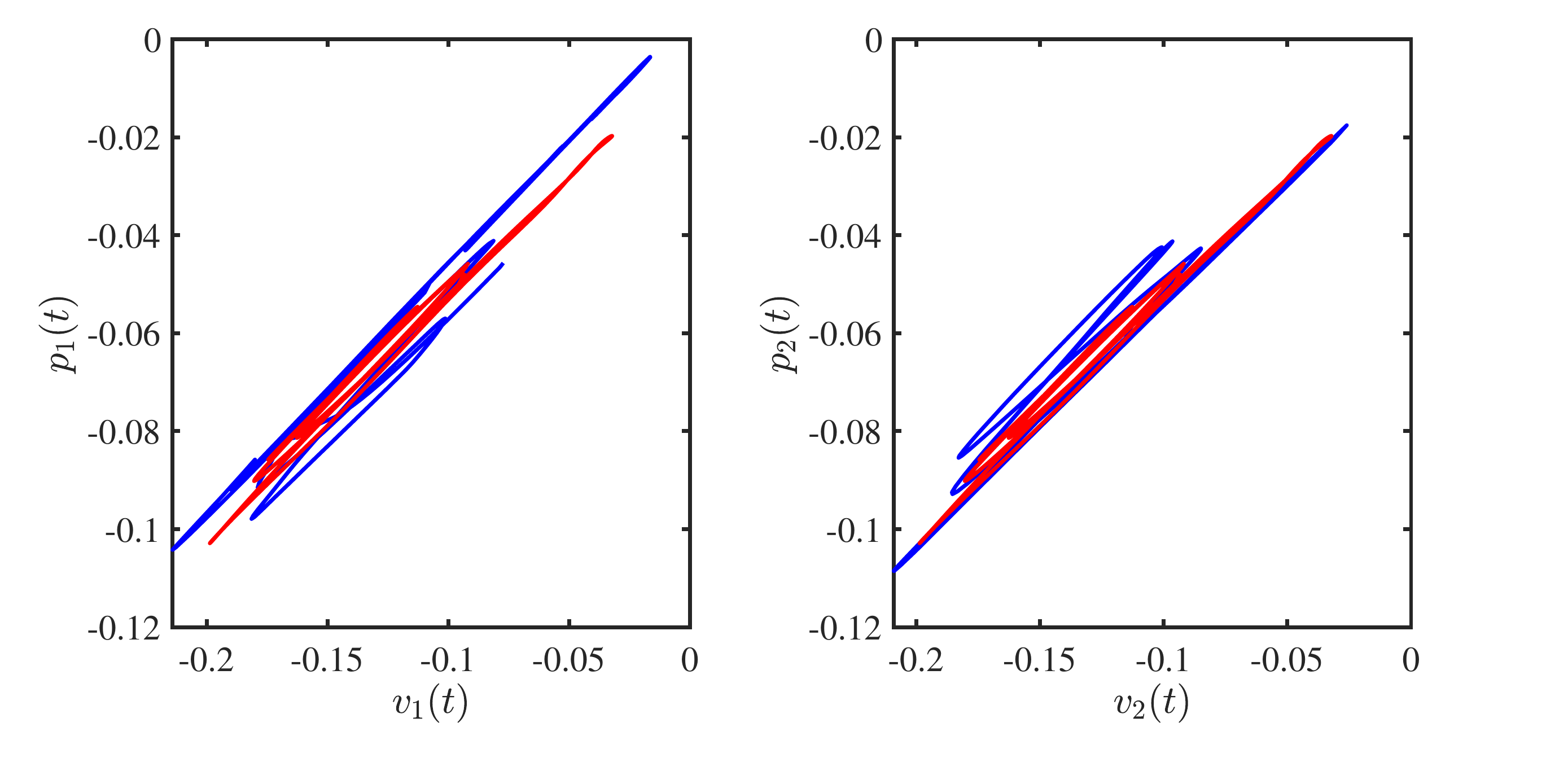}
\includegraphics[width=0.65\columnwidth]{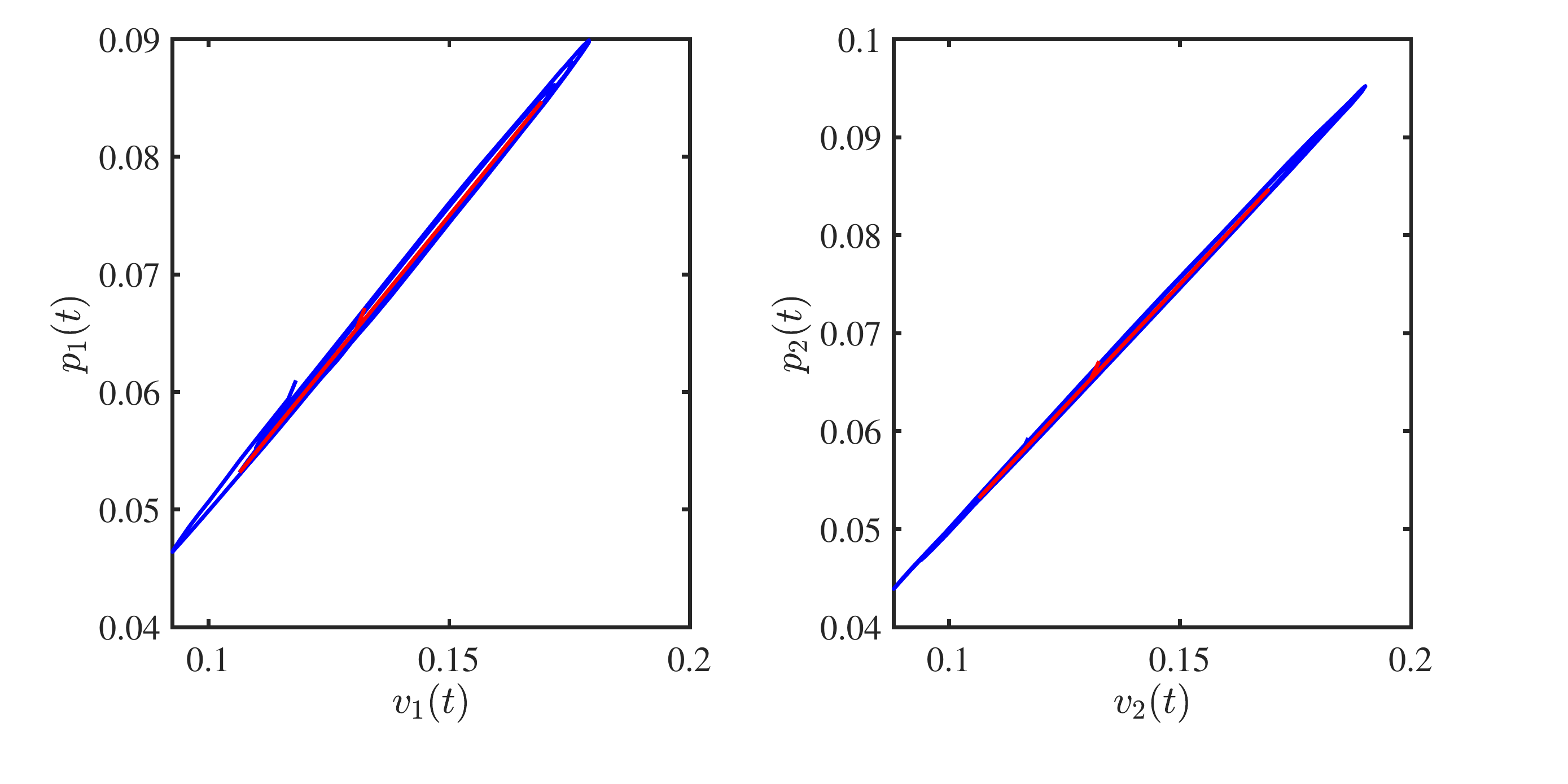}
\includegraphics[width=0.65\columnwidth]{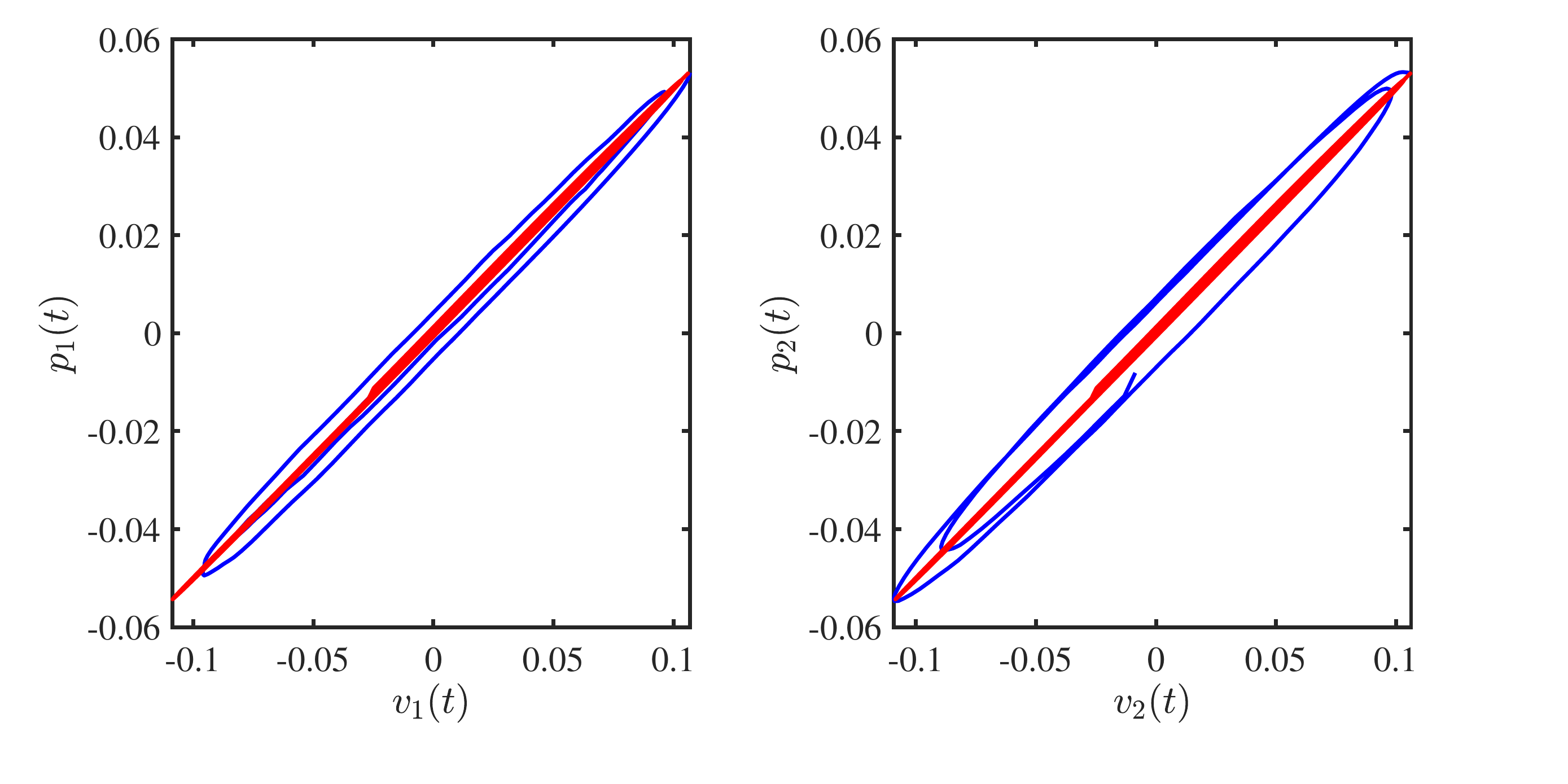}
\caption{\label{pvplots}
Plots of $p(v)$.  Case 1 (top row), Case 2 (second row), Case 4 (third row), 
and Case 5 (fourth row).  In many cases, the change of slope is not visible.
The red lines are the 8CC results whereas the blue lines are the PDE results.
Even though the 8CC  (red) curves look linear in some cases, in fact (under 
detailed examination) they are not and indicate instability.}
\end{figure}

In references \cite{PhysRevE.84.026614,PhysRevE.81.016608,PhysRevE.82.016606} 
it was shown that the stability of a solitary wave subjected to external forces 
could be inferred from the solution of the CC equations by studying the stability 
curve $p(v)$, where $p(t)$ is the momentum conjugate to $q(t)$ and $v(t) \equiv \dot{q}(t)$. 
A positive slope of the $p$ vs $v$ curve is a necessary condition for the stability 
of the solitary wave. If a branch of the $p(v)$ curve has a negative slope, this is 
a sufficient condition for instability. In our simulations, we will show that this 
criterion is consistent with the numerical simulations (see Fig.~\ref{pvplots}). 
Note that in the present setup, exact solutions are no longer available once we 
add the external potential, and simultaneously, the CC equations do not possess 
exact solutions of the form $q(t) = q_0 + v_s \, t$,  $\beta(t) = \beta_0$, $p(t) = p_0$, 
and $\theta_i(t) = \theta_{0,i} + \gamma_i \, t$. Because of this, we cannot perform 
a phase portrait analysis for solutions which are near these fixed-point solutions 
of the CC equations. Nevertheless, for most of the cases where instabilities occur, 
$p(v)$ is a good indicator of instability.
Indeed, 
we show four cases where this turnaround 
is clearly visible both for the trapped as well as moving soliton.  It is only when 
$a_2=0$ (case 3) that  we did not detect a place where $dp/dv < 0$ in our CC evolutions. 
When we increase the value of $|a_2|$ to be greater than $1/300$ the turnaround of 
the curve is much more visible than at $a_2= - 1/1000$.  We have included the numerically
determined curves $p_i(v_i)$ which show this turnaround more dramatically at $a_2= - 1/1000$.

%
%
%
%
%

%
\section{\label{s:conclusions}Conclusions}

To understand the difference between the effect of $\PT$-symmetric vs
$\PT$-antisymmetric external potentials on solitary wave dynamics, the
present work generalized the $\PT$ symmetric external potential problem studied 
in~\cite{PhysRevE.94.032213}, to a two-component NLSE in a $\PT$-antisymmetric
external potential. Imposing anti-$\PT$ symmetry on the exact solution in
the absence of the external potential requires that the masses of the two 
components are equal, and that the phases differ by $\pi/2$. 
Depending on initial conditions, the real external periodic potential can 
trap 
the solitary wave. 
On the other hand, 
the two imaginary 
anti-$\PT$ external potentials affect the solitary wave differently. 
In particular, the potential term proportional to $\sigma_3$ causes $M_1$ and $M_2$ to 
initially move in opposite directions.
Which way the masses diverge 
depends on the sign of $a_2 \cos k_2  q(0)$. The term in the potential proportional
to $a_3$  becomes more important when the term proportional to $\sigma_3$ causes
$M_1$ to differ from $M_2$. Then it tends to accelerate the collapse of one 
component and/or accelerate the blowup or collapse of the entire soliton. Otherwise, 
when $a_2=0$, the effect of $a_3$ on the behavior of the soliton is initially quite small and is negligible in the CC approximation. 
However, the numerical simulations show that eventually the presence of $a_3$ leads to an instability. 
When $a_{2}\geq 1/1000$, we observe that the instability criterion determined
by $dp/dv \leq 0$ is being met, and can be seen visually either in the 8CC 
approximation or the numerical solution of the moment equations.
We displayed cases where $M_1+M_2$ and $\beta(t)$ get larger and larger signaling 
blowup. Even in the trapped cases, when $a_2$ is still quite small, the $p(v)$ 
criterion predicts dynamic instability which is seen in the simulations. 

In all cases the lower order moments, $M_1(t)$, $M_2(t)$, $q(t)$, and $p(t)$ are 
well described by the CC equations whereas $\beta(t)$ and $\Lambda(t)$ are just 
qualitatively in agreement with the numerical solution of the PDEs. The values 
$g=2$ and $\beta(0)=1/2$ were chosen to compare our results with the $\PT$-symmetric 
single-component NLSE results. Because of the destabilizing effect of $\sigma_2$, 
there is in general no small oscillation theory for the anti-$\PT$ external potential 
problem. This is a 
major difference from the $\PT$-symmetric one-component NLSE. Whenever $a_2 \neq 0$
holds, one finds dynamic instability which explains why we were unable to find stable 
solutions of the anti-$\PT$ symmetric NLSE in the presence of an anti-$\PT$ symmetric 
external potential. We considered cases in this paper where the soliton was trapped by 
the real potential $V_0(x)$ as well as cases where the solitary wave was moving. In both
cases the 8CC approximation gave a reasonable description of the motion of the two 
components of the wave function. The phase space of possible behaviors is huge, and we 
reported on a few representative cases.
 
This work paves the way for future directions of study. At the level of the NLSEs, a 
systematic stability analysis around the steady-state and moving soliton solutions over
$(a_{1},a_{2},a_{3})$ will identify potential intervals of stability 
of 
the pertinent waveforms. If the solutions obtained are identified as 
unstable, 
then it would be interesting to corroborate even further our dynamical instability criterion 
employed in this work, i.e., $dp/dv<0$. Also, another direction of future work involves 
other kind of external potentials, such as hyperbolic ones in the form of 
$V_1(x)=\sech\,x\,\tanh{x}$ and $V_2(x) = \rmi W(x) = \rmi\sech^2 {x}$. Those directions are currently under consideration and results
will be reported in future publications.

%
\ack
EGC, FC and JFD would like to thank the Santa Fe Institute and the Center 
for Nonlinear Studies at Los Alamos National Laboratory for their hospitality.
AK is grateful to Indian National Science Academy (INSA) for awarding him 
INSA Senior Scientist position at Savitribai Phule Pune University, Pune, 
India. The work of AS was supported by the U.S.\ Department of Energy. 
%
\appendix
%
\section{\label{s:8CCderivation}Derivation of the eight component CC
equations of motion}

%
\subsection{\label{ss:Dynamic}Dynamic term}

>From Eq.\ef{Tdef}, the dynamic term splits into the sum of two 
independent parts:
\begin{equation}\label{Ttot1t2}
   T(Q,\dot{Q})
   =
   t_1(Q,\dot{Q}) + t_2(Q,\dot{Q})
   =
   \pi_{\mu}(Q) \, \dot{Q}^{\mu} \>,
   \notag
\end{equation}
where
\begin{subeqnarray}\label{Tcalc}
   t_1(Q,\dot{Q})
   &=
   M_1 \,
   \Biggl \{\,
      \dot{\theta}_1 + p \, \dot{q} -  \frac{\pi^2}{12\,\beta^2} \, \dot{\Lambda} \,
   \Biggr \} \>,
   \label{Tcalc-a} \\
   t_2(Q,\dot{Q})
   &=
   M_2 \,
   \Biggl \{\,
      \dot{\theta}_2 + p \, \dot{q} -  \frac{\pi^2}{12\,\beta^2} \, \dot{\Lambda} \,
   \Biggr \} \>,
   \label{Tcalc-b}   
\end{subeqnarray}   
so that
\begin{equation}\label{pidefs}
   \fl
   \pi_{\theta_1} = M_1
   \qc
   \pi_{\theta_2} = M_2
   \qc
   \pi_{q} = (M_1 + M_2) \, p
   \qc
   \pi_{\Lambda} = - \frac{\pi^2}{12\,\beta^2} \, (M_1 + M_2) \>.
\end{equation}
From these 
expressions, the symplectic matrix is:
\begin{eqnarray}\label{fmunu}
   \fl
   f_{\mu \nu}(Q) 
   = 
   \partial_\mu \pi_\nu- \partial_\nu \pi_\mu 
   =
   \left ( \begin{array}{cccccccc}
      0 & 1 & 0 & 0 & p & 0 & 0 & -c \\
      -1 & 0 & 0 & 0 & 0 & 0 & 0 & 0 \\
      0 & 0 & 0 & 1 & p & 0 & 0 & -c \\
      0 & 0 & -1 & 0 & 0 & 0 & 0 & 0 \\
      -p & 0 & -p & 0 & 0 & -M & 0 & 0 \\
      0 & 0 & 0 & 0 & M & 0 & 0 & 0 \\
      0 & 0 & 0 & 0 & 0 & 0 & 0 & M d \\
      c & 0 & c & 0 & 0 & 0 & -M d & 0 
   \end{array} \right ) \>,
\end{eqnarray}
where
\begin{equation}\label{Mcd}
   M = M_1 + M_2
   \qc
   c = \frac{\pi^2}{12\,\beta^2}
   \qc
   d = \frac{\pi^2}{6\,\beta^3} \>.
\end{equation}
The determinant of $f_{\mu \nu}(Q)$ is $d^2 M^4$, 
and its inverse is given by:
\begin{equation}\label{finverse}
   f^{\mu\nu}(Q)
   =
   \frac{1}{M}
   \left ( \begin{array}{cccccccc}
      0 & -M & 0 & 0 & 0 & 0 & 0 & 0 \\
      M & 0 & 0 & 0 & 0 & -p & c/d & 0 \\
      0 & 0 & 0 & -M & 0 & 0 & 0 & 0 \\
      0 & 0 & M & 0 & 0 & -p & c/d & 0 \\
      0 & 0 & 0 & 0 & 0 & 1 & 0 & 0 \\
      0 & p & 0 & p & -1 & 0 & 0 & 0 \\
      0 & -c/d & 0 & -c/d & 0 & 0 & 0 & -1/d \\
      0 & 0 & 0 & 0 & 0 & 0 & 1/d & 0 
   \end{array} \right ) \>,
\end{equation}
where
\begin{equation}\label{dcd}
   \frac{1}{d} = \frac{6\,\beta^3}{\pi^2}
   \qc
   \frac{c}{d} = \frac{\beta}{2} \>.
\end{equation}

%
\subsection{Hamiltonian and its decomposition}
%

Based on Eq.~\ef{e:VT-4}, the Hamiltonian can be written as the sum of 
three parts:
\begin{equation}\label{HamThreeParts}
   H(Q) = H_{\mathrm{kin}}(Q) + H_{\mathrm{pot}}(Q) + H_{\mathrm{nl}}(Q) \>,
\end{equation}
where $H_{\mathrm{kin}}$, $H_{\mathrm{pot}}$, $H_{\mathrm{nl}}$ stand for 
the kinetic, potential, and nonlinear terms, respectively. 

Let us consider the kinetic term first. Using the integral definitions 
of \ref{s:Integrals}, we find:
\begin{eqnarray}\label{Hkin}
\fl
   H_{\mathrm{kin}}(Q)
   =
   \tint \dd{x} \,|\partial_x \Psi_i(x,Q) |^2 
   =
   M \,
   \Biggl \{
      p^2
      +
      \frac{1}{3} \,\beta^2
      +
      \frac{\pi^2}{3} \, \frac{\Lambda^2}{\beta^2} \,
   \Biggr \} \>. 
\end{eqnarray}
In a similar fashion, the potential term gives
\begin{eqnarray}
   \fl
   H_{\mathrm{pot}}(Q)
   &=
   \tint  \dd{x} \, V_0(x) \, |\Psi(x,t)|^2
   =
   [\, A_1^2 \, \beta + A_2^2 \, \beta ] \, \beta \, a_1 \tint \dd{x}
   \sech^2[\beta (x - q)] \, \cos{k_1 x}
   \notag \\
   \fl
   &=
   \frac{M}{2} \, a_1 \cos(k_1 q) \, G_1(k_1/\beta) \>,
   \label{Hpot}
\end{eqnarray}
where $G_1(z)$ is given in \ef{Gints-a}.
Finally, we consider the nonlinear term,
\begin{eqnarray}
   \fl
   H_{\mathrm{nl}}(Q) 
   &=
   - \frac{g}{2}
   \tint dx \,|\Psi(x,Q)|^{4}
   =
   - \frac{g}{6} \, \beta \, M^2 \>.
   \label{Hnl}
\end{eqnarray}
%
From \ef{Hkin}, \ef{Hpot}, and \ef{Hnl}, the Hamiltonian is given by
\begin{equation}\label{Hfull}
   \fl
   H(Q)
   =
   M \,
   \Biggl \{
      p^2
      +
      \frac{1}{3} \,\beta^2
      +
      \frac{\pi^2}{3} \, \frac{\Lambda^2}{\beta^2}
      +
      \frac{a_1}{2} \, \cos(k_1 q) \, G_1(k_1/\beta) \,
   \Biggr \} 
   -
   \frac{g}{6} \, \beta \, M^2 \>.
\end{equation}
Defining
\begin{equation} \label{vmu2}
   v_{\mu}(Q) = \partial_{\mu} H(Q) \>,
\end{equation} 
the nonzero derivatives of the Hamiltonian with respect to the parameters are given by:
\begin{subeqnarray}\label{vmu}
   \fl
   v_{M_1}
   &=v_{M_2} =
   p^2
   +
   \frac{1}{3} \,\beta^2
   +
   \frac{\pi^2}{3} \, \frac{\Lambda^2}{\beta^2}
   +
   \frac{a_1}{2} \, \cos(k_1 q) \, G_1(k_1/\beta)
   -
   \frac{g}{3} \, \beta \, M \>,
   \label{vM1} \\
   \fl
   v_{q}
   &=
   - M \, \frac{k_1 a_1}{2} \, \sin(k_1 q) \, G_1(k_1/\beta) \>,
   \label{vq} \\
   \fl
   v_{p}
   &=
  2 \,  M \, p \>,
   \label{vp} \\
   \fl
   v_{\beta}
   &=
   M \,
   \Biggl \{
      \frac{2}{3} \,\beta
      -
      \frac{2 \pi^2}{3} \, \frac{\Lambda^2}{\beta^3}
      -
      \frac{a_1 k_1}{2 \beta^2} \, \cos( k_1 q) \, G'_1(k_1/\beta) \,
   \Biggr \}
   -
   \frac{g}{6} \, M^2 \>,
   \label{vbeta} \\
   \fl
   v_{\Lambda}
   &=
   M \, \frac{2 \pi^2}{3} \, \frac{\Lambda}{\beta^2} \>.
   \label{vLambda}
\end{subeqnarray}
Here $G'_1(z)$ is given in \ef{Gints-d}.

%
\subsection{\label{ss:disterm}Dissipative term}

From \ef{U0U1def} and \ef{e:VT-4.1}, the dissipative term splits into two parts.  We find
\begin{eqnarray}
   \fl
   F(Q,\dot{Q})
   &=
   \rmi \tint \dd{x} 
   \bigl \{\,
      \Psi^{\dag}(x,Q) \, U_1(x) \, \Psi_t(x,Q)
      -
      \Psi_t^{\dag}(x,Q) \, U_1(x) \Psi(x,Q) \,
   \bigr \}
   \notag \\
   \fl
   &=
   F_1(Q,\dot{Q}) + F_2(Q,\dot{Q}) \>,
   \label{diss0}
\end{eqnarray}
where
\begin{subeqnarray}\label{Fij}           
   F_1(Q,\dot{Q})
   &=
   a_2 \, [\, F_{11}(Q,\dot{Q}) - F_{22}(Q,\dot{Q}) \, ] \>,
   \label{F1122} \\
   F_2(Q,\dot{Q})
   &=
   a_3 \, [\, F_{12}(Q,\dot{Q}) + F_{21}(Q,\dot{Q}) \, ] \>,
   \label{F1221}
\end{subeqnarray}
with
\begin{subeqnarray}\label{Fall}
   F_{ii}(Q,\dot{Q})
   &=
   - 2 \tint \dd{x} \cos(k_2 x) 
   \Imag{ \psi^{\ast}_i(x,Q) \, \partial_t \psi_i(x,Q) }\>,
   \label{Fall-a} \\
   F_{12}(Q,\dot{Q})
   &=
   - 2 \tint \dd{x} \cos(k_3 x)
   \Imag{ \psi^{\ast}_1(x,Q) \, \partial_t \psi_2(x,Q) }\>,
   \label{Fall-b} \\
   F_{21}(Q,\dot{Q})
   &=
   - 2 \tint \dd{x} \cos(k_3 x)
   \Imag{ \psi^{\ast}_2(x,Q) \, \partial_t \psi_1(x,Q) }\>.
\end{subeqnarray}
Changing variables to $y = \beta (x - q)$, we find
\begin{eqnarray}
   \fl
   F_{ii}(Q,\dot{Q})
   &=
   M_i 
   \tint \dd{y} 
   \cos\Bigl [\, k_2 \Bigl( \frac{y}{\beta} + q \Bigr ) \,\Bigr] 
   \Bigl \{\,
      \dot{\theta}_i 
      - 
      \frac{\dot{p}}{\beta} \, y 
      + 
      p \, \dot{q}
      - 
      \frac{\dot{\Lambda}}{\beta^2} \, y^2
      +
      \frac{2 \Lambda \dot{q}}{\beta} \, y \,
   \Bigr \}
   \sech^2(y)
   \notag \\
   \fl
   &=
   M_i 
   \bigl \{ \,
      \cos(k_2 q) \, 
      [\,
         (\, \dot{\theta}_i + p \, \dot{q} \, ) \, G_1(k_2/\beta)
         -
         (\dot{\Lambda}/\beta^2) \, G_2(k_2/\beta) \,
      ]
      \notag \\
      \fl
      & \hspace{1.2em}
      +
      \sin(k_2 q) \,
      [\, (\, \dot{p} - 2 \Lambda \dot{q} \,)/ \beta \,] \,
      G_3(k_2/\beta) \,
   \bigr \} \>.
   \label{F1x}
\end{eqnarray}
So from \ef{F1122}, we find
\begin{eqnarray}
   \fl
   F_1(Q,\dot{Q})
   &=
   a_2 \, M_1 
   \bigl \{ \,
      \cos(k_2 q) \, 
      [\,
         (\, \dot{\theta}_1 + p \, \dot{q} \, ) \, G_1(k_2/\beta)
         -
         (\dot{\Lambda}/\beta^2) \, G_2(k_2/\beta) \,
      ]
      \notag \\
      \fl
      & \hspace{1em}
      +
      \sin(k_2 q) \,
      [\, (\, \dot{p} - 2 \Lambda \dot{q} \,)/ \beta \,] \,
      G_3(k_2/\beta) \,
   \bigr \} 
   \notag \\
   \fl
   & \hspace{1em}
   -
   a_2 \, M_2 
   \bigl \{ \,
      \cos(k_2 q) \, 
      [\,
         (\, \dot{\theta}_2 + p \, \dot{q} \, ) \, G_1(k_2/\beta)
         -
         (\dot{\Lambda}/\beta^2) \, G_2(k_2/\beta) \,
      ]
      \notag \\
      \fl
      & \hspace{1em}
      +
      \sin(k_2 q) \,
      [\, (\, \dot{p} - 2 \Lambda \dot{q} \,)/ \beta \,] \,
      G_3(k_2/\beta) \,
   \bigr \}  \>.
   \label{F1QdotQ}
\end{eqnarray}
Defining
\begin{equation}\label{wdefs1}
   w_{1,\mu}(Q) = \pdv{F_1(Q,\dot{Q})}{\dot{Q}^{\mu}} \>,
\end{equation}
we find the non-zero components are:
\begin{subeqnarray}\label{w1values}
  \fl
   w_{1,\theta_1}
   &=
   a_2 \, M_1 \, \cos(k_2 q) \, G_1(k_2/\beta) \>,
   \label{w1-b} \\
   \fl
   w_{1,\theta_2}
   &=
   -
   a_2 \, M_2 \, \cos(k_2 q) \, G_1(k_2/\beta) \>,
   \label{w1-d} \\
   \fl
   w_{1,q}
   &=
   a_2 \, (M_1 - M_2) \,
   \Bigl [\, 
      p \cos(k_2 q) \, G_1(k_2/\beta) 
      - 
      \frac{2 \Lambda}{\beta} \sin(k_2 q) \, G_3(k_2/\beta) \,
   \Bigr ] \>,
   \label{w1-e} \\
   \fl
   w_{1,p}
   &=
   a_2 \, \frac{M_1 - M_2}{\beta} \sin(k_2 q) \, G_3(k_2/\beta) \>,
   \label{w1-f} \\
   \fl
   w_{1,\Lambda}
   &=
   - a_2 \, \frac{M_1 - M_2}{\beta^2} \cos(k_2 q) \, G_2(k_2/\beta) \>.
   \label{w1-h}
\end{subeqnarray}
For $F_{12}(Q,\dot{Q})$ and again setting $y = \beta (x - q)$,  we  find
\begin{eqnarray}\label{need3}
   \fl
   F_{12}(Q,\dot{Q})
   &=
   - 2 \tint \dd{x} \cos(k_3 x)
   \Imag{ \psi_1^{\ast}(x,t) \, [\, \partial_t \psi_2(x,t) \,] }
   \\
   \fl
   &=
   \sqrt{M_1 M_2}
   \tint \dd{y} \cos[\, k_3 (y / \beta) + k_3 q \,]\, \sech^2(y) \,
   \notag \\[-5pt]
   \fl
   & \hspace{5em}
   \times
   \bigl \{\,
      \cos(\theta_1 - \theta_2) \,
      \bigl [\,
         -
         \dot{p} y / \beta
         +
         p \dot{q}
         -
         \dot{\Lambda} y^2 / \beta^2
         +
         2 \Lambda \dot{q} y / \beta
         +
         \dot{\theta}_2 \, 
      \bigr ] \,
   \notag \\
   \fl
   & \hspace{6em}
      -
      \sin(\theta_1 - \theta_2) \,
      \bigl [\,
         \dot{A}_2 / A_2 + \dot{\beta}/\beta 
         - 
         ( \dot{\beta} y / \beta - \beta \dot{q} ) \, \tanh(y) \,
      \bigr ] \,
   \bigr \} \>.
   \notag
\end{eqnarray}
Simplifying, we obtain
\begin{eqnarray}\label{F12}
   \fl
   F_{12}(Q,\dot{Q})
   &= 
   \sqrt{M_1 M_2} \,
   \\
   \fl
   & \hspace{0em}
   \times
   \bigl \{\,
      \cos(k_3 q) \, \cos(\theta_1 - \theta_2) \,
      \bigl [\,
         (\, \dot{\theta}_2 + p\, \dot{q} \,) \, G_1(z) 
         - 
         ( \dot{\Lambda}/\beta^2) \, G_2(z) \,
      \bigr ]
      \notag \\
      \fl
      & \hspace{0.7em}
      -
      \cos(k_3 q) \, \sin(\theta_1 - \theta_2) \,
      \bigl [\,
         \dot{M}_2 / M_2 + \dot{\beta}/\beta \,] G_1(z) /2 
         +
         (\dot{\beta}/\beta) \, G_5(z)\,
      \bigr ]
      \notag \\
      \fl
      & \hspace{0.7em}
      +
      \sin(k_3 q) \, \cos(\theta_1 - \theta_2) \,
      \bigl [\, 
         (\, \dot{p} - 2 \Lambda \dot{q} \,) \, G_3(z) /\beta \,
      \bigr ]
      \notag \\
      \fl
      & \hspace{0.7em}
      +
      \sin(k_3 q) \, \sin(\theta_1 - \theta_2) \,
      [\, \beta \dot{q} \, G_4(z) \, ] \,
   \bigr \} \>.
   \notag
\end{eqnarray}
Similarly,
\begin{eqnarray}\label{F21}
   \fl
   F_{21}(Q,\dot{Q})
   &= 
   \sqrt{M_1 M_2} \,
   \\
   \fl
   & \hspace{0em}
   \times
   \bigl \{\,
      \cos(k_3 q) \, \cos(\theta_1 - \theta_2) \,
      \bigl [\,
         (\, \dot{\theta}_1 + p\, \dot{q} \,) \, G_1(z) 
         - 
         ( \dot{\Lambda}/\beta^2) \, G_2(z) \,
      \bigr ]
      \notag \\
      \fl
      & \hspace{0.7em}
      +
      \cos(k_3 q) \, \sin(\theta_1 - \theta_2) \,
      \bigl [\,
         \dot{M}_1 / M_1 + \dot{\beta}/\beta \,] G_1(z) /2 
         +
         (\dot{\beta}/\beta) \, G_5(z)\,
      \bigr ]
      \notag \\
      \fl
      & \hspace{0.7em}
      +
      \sin(k_3 q) \, \cos(\theta_1 - \theta_2) \,
      \bigl [\, 
         (\, \dot{p} - 2 \Lambda \dot{q} \,) \, G_3(z) /\beta \,
      \bigr ]
      \notag \\
      \fl
      & \hspace{0.7em}
      -
      \sin(k_3 q) \, \sin(\theta_1 - \theta_2) \,
      [\, \beta \dot{q} \, G_4(z) \, ] \,
   \bigr \} \>.
   \notag
\end{eqnarray}
So from \ef{F1221}, combining \ef{F12} and \ef{F21}, we get
\begin{eqnarray}\label{e:F2}
   \fl
   F_2(Q,\dot{Q})
   &=
   a_3 \, [\, F_{12}(Q,\dot{Q}) + F_{21}(Q,\dot{Q}) \, ] \>
   \notag \\
   \fl
   &=
   a_3 \, \sqrt{M_1 M_2} \,
   \\
   \fl
   & \hspace{0em}
   \times
   \bigl \{\,
      \cos(k_3 q) \, \cos(\theta_1 - \theta_2) \,
      \bigl [\,
         (\, \dot{\theta}_1 + \dot{\theta}_2 + 2\, p\, \dot{q} \,) \, G_1(z) 
         - 
         2 \, ( \dot{\Lambda}/\beta^2) \, G_2(z) \,
      \bigr ]
      \notag \\
      \fl
      & \hspace{0.7em}
      +
      \cos(k_3 q) \, \sin(\theta_1 - \theta_2) \,
      \bigl [\,
         (\, \dot{M}_1 / M_1 - \dot{M}_2 / M_2 \,)\, G_1(z) /2 
      \bigr ]
      \notag \\
      \fl
      & \hspace{0.7em}
      +
      \sin(k_3 q) \, \cos(\theta_1 - \theta_2) \,
      \bigl [\, 
         2 \, (\, \dot{p} - 2 \Lambda \dot{q} \,) \, G_3(z) /\beta \,
      \bigr ] \,
   \bigr \} \>.
   \notag
\end{eqnarray}
Defining
\begin{equation}\label{wdefs2}
   w_{2,\mu}(Q) = \pdv{F_2(Q,\dot{Q})}{\dot{Q}^{\mu}} \>,
\end{equation}
we find
\begin{subeqnarray}\label{w2values}
   \fl
   w_{2,M_1}
   &=
   a_3 \sqrt{M_2/M_1} \, \cos( k_3 q) \, \sin(\theta_1 - \theta_2) \,
   G_1(k_3/\beta) / 2 \>,
   \label{w2-a} \\
   \fl
   w_{2,\theta_1}
   &=
   a_3 \sqrt{M_1 M_2} \, \cos( k_3 q) \, \cos(\theta_1 - \theta_2) \,
   G_1(k_3/\beta) \>,
   \label{w2-b} \\
   \fl
   w_{2,M_2}
   &=
   - a_3 \sqrt{M_1/M_2} \, \cos( k_3 q) \, \sin(\theta_1 - \theta_2) \,
   G_1(k_3/\beta) / 2 \>,
   \label{w2-c} \\
   \fl
   w_{2,\theta_2}
   &=
   a_3 \sqrt{M_1 M_2} \, \cos( k_3 q) \, \cos(\theta_1 - \theta_2) \,
   G_1(k_3/\beta) \>,   
   \label{w2-d} \\
   \fl
   w_{2,q}
   &=
   a_3 \sqrt{M_1 M_2} \,
   \bigl \{\,
      \cos( k_3 q) \, \cos(\theta_1 - \theta_2) \,  2 \, p \, G_1(k_3/\beta)
      \notag \\
      \fl
      & \hspace{6em}
      -
      \sin(k_3 q) \, \cos(\theta_1 - \theta_2) \, 
      (4 \, \Lambda/\beta) \, G_3(k_3/\beta)
   \bigr \} \>, 
   \label{w2-e} \\
   \fl
   w_{2,p}
   &=
   a_3 \sqrt{M_1 M_2} \, \sin(k_3 q) \, 
   \cos(\theta_1 - \theta_2) \, (2/\beta) \, G_3(k_3/\beta) \>,
   \label{w2-f} \\
   \fl
   w_{2,\beta}
   &=
   0 \>,
   \label{w2-g} \\
   \fl
   w_{2,\Lambda}
   &=
   - a_3 \sqrt{M_1 M_2} \, \cos(k_3 q) \, \cos(\theta_1 - \theta_2) \,
   (2/\beta^2) \, G_2(k_3/\beta) \>.
   \label{w2-h}
\end{subeqnarray}

%
\subsection{\label{ss:varEOM}Equations of motion}

From \ef{e:VT-9}, the equations of motion are found from
\begin{equation}\label{}
   \dot{Q}^\mu
   =
   f^{\mu\nu}(Q) \, u_{\mu}(Q) 
   \qc
   u_{\mu}(Q) = v_{\mu}(Q) - w_{\mu}(Q) \>.   
\end{equation}
Let us first find $u_{\mu}(Q)$.  From \ef{vmu}, \ef{w1values}, and \ef{w2values},
\begin{subeqnarray}\label{umu}
   \fl
   u_{M_1}
   &=
   p^2
   +
   \frac{1}{3} \,\beta^2
   +
   \frac{\pi^2}{3} \, \frac{\Lambda^2}{\beta^2}
   +
   \frac{a_1}{2} \, \cos(k_1 q) \, G_1(k_1/\beta)
   -
   \frac{g}{3} \, \beta \, M \, 
   \label{uM1} \\
   \fl
   & \hspace{2em}
   -
   a_3 \sqrt{M_2/M_1} \, \cos( k_3 q) \, \sin(\theta_1 - \theta_2) \,
   G_1(k_3/\beta) / 2 \>,
   \notag \\
   \fl
   u_{\theta_1} 
   &=
   - a_2 \, M_1 \, \cos(k_2 q) \, G_1(k_2/\beta)
   \label{utheta1} \\
   \fl
   & \hspace{2em}
   -
   a_3 \sqrt{M_1 M_2} \, \cos( k_3 q) \, \cos(\theta_1 - \theta_2) \,
   G_1(k_3/\beta) \>,
   \notag \\
   \fl
   u_{M_2}
   &=
   p^2
   +
   \frac{1}{3} \,\beta^2
   +
   \frac{\pi^2}{3} \, \frac{\Lambda^2}{\beta^2}
   +
   \frac{a_1}{2} \, \cos(k_1 q) \, G_1(k_1/\beta)
   -
   \frac{g}{3} \, \beta \, M 
   \label{uM2} \\
   \fl
   & \hspace{2em}
   +
   a_3 \sqrt{M_1/M_2} \, \cos( k_3 q) \, \sin(\theta_1 - \theta_2) \,
   G_1(k_3/\beta) / 2 \>,
   \notag \\
   \fl
   u_{\theta_2} 
   &=
   a_2 \, M_2 \, \cos(k_2 q) \, G_1(k_2/\beta)
   \label{utheta2} \\
   \fl
   & \hspace{2em}
   -
   a_3 \sqrt{M_1 M_2} \, \cos( k_3 q) \, \cos(\theta_1 - \theta_2) \,
   G_1(k_3/\beta) \>,   
   \notag \\
   \fl
   u_{q}
   &=
   - M \, \frac{k_1 a_1}{2} \, \sin(k_1 q) \, G_1(k_1/\beta)
   \notag \\
   \fl
   & \hspace{2em}
   -
   a_2 \, (M_1 - M_2) \,
   \bigl [\, 
      p \cos(k_2 q) \, G_1(k_2/\beta) 
      - 
      (2 \Lambda/\beta) \sin(k_2 q) \, G_3(k_2/\beta) \,
   \bigr ] 
   \label{uq} \\
   \fl
   & \hspace{2em}
   -
   a_3 \sqrt{M_1 M_2} \,
   \bigl \{\,
      \cos( k_3 q) \, \cos(\theta_1 - \theta_2) \,  2 \, p \, G_1(k_3/\beta)
      \notag \\
      \fl
      & \hspace{6em}
      -
      \sin(k_3 q) \, \cos(\theta_1 - \theta_2) \, 
      (4 \, \Lambda/\beta) \, G_3(k_3/\beta)
   \bigr \} \>, 
   \notag \\
   \fl
   u_{p}
   &=
   M \, 2 \, p
   -
   a_2 \, \frac{M_1 - M_2}{\beta} \sin(k_2 q) \, G_3(k_2/\beta) \> 
   \label{up} \\
   \fl
   & \hspace{2em}
   -
   a_3 \sqrt{M_1 M_2} \, \sin(k_3 q) \, 
   \cos(\theta_1 - \theta_2) \, (2/\beta) \, G_3(k_3/\beta) \>,
   \notag \\
   \fl
   u_{\beta}
   &=
   M \,
   \Biggl \{
      \frac{2}{3} \,\beta
      -
      \frac{2 \pi^2}{3} \, \frac{\Lambda^2}{\beta^3}
      -
      \frac{a_1 k_1}{2 \beta^2} \, \cos( k_1 q) \, G'_1(k_1/\beta) \,
   \Biggr \}
   -
   \frac{g}{6} \, M^2 \>,
   \label{ubeta} \\
   \fl
   u_{\Lambda}
   &=
   M \, \frac{2 \pi^2}{3} \, \frac{\Lambda}{\beta^2}
   +
   a_2 \, \frac{M_1 - M_2}{\beta^2} \cos(k_2 q) \, G_2(k_2/\beta)
   \label{uLambda} \\
   \fl
   & \hspace{2em}
   +
   a_3 \sqrt{M_1 M_2} \, \cos(k_3 q) \, \cos(\theta_1 - \theta_2) \,
   (2/\beta^2) \, G_2(k_3/\beta) \>.
   \notag 
\end{subeqnarray}
Using \ef{finverse} and \ef{umu} and \ef{e:VT-9}, we obtain the 8CC 
equations of motion as given by Eqs.~\ef{EOMX}.
\section{\label{s:Integrals}Useful integrals and definitions}

We 
define the following integrals:
\begin{subeqnarray}\label{Gints}
   G_1(z)
   &:=
   \tint \dd{y}
   \cos(z y)
   \sech^2(y)
   =
   \pi z \csch( \pi z / 2 ) \>,
   \label{Gints-a} \\
   G_2(z)
   &:=
   \tint \dd{y} y^2 
   \cos(z y)
   \sech^2(y)
   \notag \\
   &=
   - 
   \frac{\pi^2}{8} \csch^3( \pi z / 2 ) \,
   [\, \pi z \, (\, 3 + \cosh(\pi z) \,) - 4 \, \sinh(\pi z) \,] \,, 
   \label{Gints-b} \\ 
   G_3(z)
   &:=
   \tint \dd{y} y
   \sin(z y)
   \sech^2(y)
   \notag \\
   &=
   \frac{\pi}{2} \, \csch( \pi z / 2 ) \,
   [\, -2 + \pi z \, \coth( \pi z / 2 ) \, ] \>, 
   \label{Gints-c} \\
   G_4(z)
   &:=
   \tint \dd{y} \sin(z y) \sech^2(y) \tanh(y)
   \notag \\
   &=
   \frac{\pi \, z^2}{2} \csch(\pi z / 2) \>,
   \label{Gints-d} \\
   G_5(z)
   &:=
   \tint \dd{y} y \cos(z y) \sech^2(y) \tanh(y)
   \notag \\ 
   &=
   \frac{\pi z}{4} \, [\, 4 - \pi z \coth(\pi z/2) \, ] \, \csch(\pi z / 2) \>,
   \label{Gints-e} \\
   G_1'(z)
   &=
   \pi \csch( \pi z / 2 ) \, 
   [\, 1 - (\pi z / 2 ) \, \coth( \pi z / 2 ) \,] \>.
   \label{Gints-f}
\end{subeqnarray}
We note that
\begin{subeqnarray}\label{e:I-1}
   \frac{d}{dz} \sech(z)
   &=
   - \sech(z) \, \tanh(z) 
   \label{e:I-1a} \,, \\
    \frac{d}{dz} \tanh(z)  
   &=
   \sech^2(z) \>, 
   \label{e:I-1b}
\end{subeqnarray}
together with the following useful integrals:
\begin{subeqnarray}\label{e:I-2}
   \tint dz \sech^2(z)
   &=
   2 \>, 
   \label{e:I-2a} \\
   \tint dz\, \sech^4(z)
   &=
   \frac{4}{3} \>, 
   \label{e:I-2c} \\
   \tint dz \,z^2 \sech^2(z)
   &=
   \frac{\pi^2}{6} \>.
   \label{e:I-2d} 
\end{subeqnarray}

\section*{Bibliography}
%
%

%
\end{document}